\documentclass[11pt,preprint]{aastex}
\usepackage{apjfonts}

\usepackage{lscape}
\usepackage{lineno}

\usepackage{epsfig,color}
\usepackage{amssymb,amsmath}
\usepackage{mathrsfs}
\usepackage{graphicx}
\usepackage{bm}


\def\oiii{[O~{\sc iii}]}
\def\feii{Fe~{\sc ii}}
\def\Hb{H$\beta$}
\def\Ha{H$\alpha$}

\def\dmfg1{$\mathscr{\dot M}_{f_{\rm g}=1}$}
\def\dmfg{$\mathscr{\dot M}_{f_{\rm g}}$}

\def\ledf1{$L_{\rm Edd}(f_{\rm g}=1)$}

\shorttitle{Measuring the Virial Factor in SDSS DR7 AGNs} \shortauthors{Liu et al.}

\begin{document}

\title{Measuring the Virial Factor in SDSS DR7 AGNs with Redshifted \Hb\ and \Ha\ Broad Emission Lines}

\author{H. T. Liu\altaffilmark{1,2,3}, Hai-Cheng Feng\altaffilmark{1,2,3}, Sha-Sha Li\altaffilmark{1,2,3}, J. M. Bai\altaffilmark{1,2,3}, and H. Z. Li\altaffilmark{4} }

\altaffiltext{1} {Yunnan Observatories, Chinese Academy of Sciences, 396 Yangfangwang, Guandu District, Kunming 650216, Yunnan, People's Republic of China; htliu@ynao.ac.cn, hcfeng@ynao.ac.cn}

\altaffiltext{2} {Key Laboratory for the Structure and Evolution of Celestial Objects, Chinese Academy of Sciences, Kunming 650216, Yunnan, People's Republic of China}

\altaffiltext{3} {Center for Astronomical Mega-Science, Chinese Academy of Sciences, 20A Datun Road, Chaoyang District, Beijing 100012, People's Republic of China}

\altaffiltext{4} {Physics Department, Yuxi Normal University, 134 Fenghuang Road, Hongta District, Yuxi 653100, Yunnan, People's Republic of China}


\begin{abstract}
  Under the hypothesis of gravitational redshift induced by the central supermassive black hole, and based on line widths and shifts of redward shifted \Hb\ and \Ha\ broad emission lines for more than 8000 SDSS DR7 AGNs, we measure the virial factor in determining supermassive black hole masses. The virial factor had been believed to be independent of accretion radiation pressure on gas clouds in broad-line region (BLR), and only dependent on inclination effects of BLR. The virial factor measured spans a very large range. For the vast majority of AGNs ($>$96\%) in our samples, the virial factor is larger than $f=1$ usually used in literatures. The $f$ correction makes the percent of high-accreting AGNs decrease by about 100 times. There are positive correlations of $f$ with the dimensionless accretion rate and Eddington ratio. The redward shifts of \Hb\ and \Ha\ are mainly the gravitational origin, confirmed by a negative correlation between the redward shift and the dimensionless radius of BLR. Our results show that radiation pressure force is a significant contributor to the measured virial factor, containing the inclination effects of BLR. The usually used values of $f$ should be corrected for high-accreting AGNs, especially high redshift quasars. The $f$ correction increases their masses by one--two orders of magnitude, which will make it more challenging to explain the formation and growth of supermassive black holes at high redshifts.

\end{abstract}


\section{INTRODUCTION}\label{sec:intr}
Black hole mass, $M_{\bullet}$, is an important fundamental parameter of black hole. Reliable measurement of $M_{\bullet}$ always is a key issue of black hole related researches.
For active galactic nuclei (AGNs), the reverberation mapping (RM) method or the relevant
secondary methods based on single-epoch spectra were widely used to measure $M_{\bullet}$
by a virial mass $M_{\rm{RM}}=f v^2_{\rm{FWHM}} r_{\rm{BLR}}/G$ when clouds in broad-line region (BLR) are in virialized motion, where $f$ is the virial factor, $v_{\rm{FWHM}}$ is full width at half maximum of broad emission line, $r_{\rm{BLR}}$ is radius of BLR, and $G$ is the gravitational constant \citep[e.g.,][]{Pe04}. However, $f$ is very uncertain due to the unclear kinematics and geometry of BLR \citep[e.g.,][]{Pe04,Wo15}. $f$ is commonly considered to be the main source of uncertainty in $M_{\rm{RM}}$. The reverberation-based masses are themselves uncertain typically by a factor of $\sim$ 2.9 \citep{On04}, and the absolute uncertainties in $M_{\rm{RM}}$ given by the secondary methods are typically around a factor of 4 \citep{VP06}. If $v_{\rm{FWHM}}$ is replaced with the line width $\sigma_{\rm{line}}$, the second moment of emission line, $f$ becomes $f_{\sigma}$. Based on the photoionization assumption \citep[e.g.,][]{BM82,Pe93}, $r_{\rm{BLR}}= \tau_{\rm{ob}} c/(1+z)$, where $c$ is the speed of light, $z$ is the cosmological redshift of source, and $\tau_{\rm{ob}}$ is the observed time lag of the broad-line variations relative to the continuum ones. For non--RM AGNs studied by the secondary methods, $r_{\rm{BLR}}$ can be estimated with the empirical $r_{\rm{BLR}}$--$L_{\rm{5100}}$ relation for \Hb\ emission line of the RM AGNs, where $L_{\rm{5100}}$ is AGN continuum luminosity at rest-frame wavelength 5100 \AA\ \citep[e.g.,][]{Ka00,Be13,Du18b,Du19,Yu20}.

RM surveys had been \textbf{carried out} \citep[e.g.,][]{Ki15,Sh15a,Sh15b,Sh16,Gr17,Ho19,Sh19}. Non-survey RM observation researches had been made for more than 100 AGNs over the last several decades \citep[e.g.,][]{Ka99,Ka00,Pe05,Be06,Ka07,Be10,De10,Ba11,Ha11,Po12,Du14,Pe14,Wa14,Ba15,
Du15,Hu15,Be16,Du16,Lu16,Pe17,Du18a,Du18b,Xi18a,Xi18b,Zh19,Hu20,Be21,Fe21a,Fe21b,Hu21,LYY21,
Lu21,Be22,LFL22,Be23}. The single-epoch spectra had been widely used to estimate $M_{\rm{RM}}$ \textbf{in studies of} high-$z$ quasars \citep[e.g.,][]{Wi10,Wu15,Wa19,Ei23}, and on statistics of AGNs, such as the Sloan Digital Sky Survey (SDSS) quasars \citep[e.g.,][]{Hu08,Li19}. Based on the $M_{\bullet}-\sigma_{\ast}$ relation for the low-$z$ inactive and quiescent galaxies with $\sigma_{\ast}$ to be stellar velocity dispersion of galaxy bulge \citep[e.g.,][]{Tr02,On04,Pi15,Wo15}, these derived averages of $\langle f\rangle \approx 1$ and/or $\langle f_{\rm{\sigma}}\rangle \approx 5$ were usually used to estimate $M_{\rm{RM}}$ by the RM and/or single-epoch spectra of AGNs. Therefore, measuring $f$ and/or $f_{\rm{\sigma}}$ independently by a new method for individual AGNs is necessary and important to understand the physics of BLR, and the issues related to masses of supermassive black holes (SMBHs), e.g., the formation and growth of SMBHs at $z\gtrsim 6$ \citep[e.g.,][]{Wu15,Fa23}, coevolution (or not) of SMBHs and host galaxies \citep[e.g.,][]{Tr02,Ko13,Wo13,Ca20}, etc.

Some efforts have been made on an object by object basis for small samples of AGNs using high-fidelity RM techniques \citep[e.g.,][]{Pa14a,Pa14b} or by using spectral fitting methods \citep[][and references therein]{MR18}. \citet{Li17} proposed a new method to measure $f$ based on the widths and shifts of redward shifted broad emission lines for the RM AGNs. Based on SDSS DR 5 quasars with redward shifted \Hb\ and \feii\ broad emission lines, \citet{Li22} made further efforts of researching $f$ and $f_{\rm{\sigma}}$. Fe {\sc iii}$\lambda\lambda$ 2039--2113 UV line blend comes from an inner region of BLR \citep{MJ18,MJ21}, and for 10 lensed quasars of higher Eddington ratio, the redward shifted Fe {\sc iii} blend was used to estimate $f$ with $\langle f\rangle=14.3$ much larger than $\langle f\rangle \approx 1$ \citep{MJ20}. However, the origins of broad emission lines and BLR are yet unclear for AGNs \citep[e.g.,][]{Wa17}. Thus, it is unclear of the origin of redward shifts of broad emission lines. The redward shifts of broad emission lines are commonly believed to be from inflow \citep[e.g.,][]{Hu08}. Inflow can generate the redward shifts of broad absorption lines, but the broad absorption and emission lines may be from different gas regions due to their distinct velocities \citep{ZS19}.

RM observations of Mrk 817 suggest that the redward shifts of broad emission lines do not originate from inflow because of their redward asymmetric velocity-resolved lag maps \citep{Lu21}, which are not consistent with the blueward asymmetric maps expected from inflow. Redward shifts of broad emission lines in the RM observations of Mrk 110 follow the gravitational redshift prediction \citep{Ko03}. The gravitational interpretation of redward shift of the Fe {\sc iii} blend is preferred than alternative explanations, such as inflow, that will need additional physics to explain the observed correlation between the width and redward shift of the blend \citep{MJ18}. A sign of the gravitational redshift $z_{\rm{g}}$ was found in a statistical sense for broad \Hb\ in the single-epoch spectra of SDSS DR 7 quasars \citep{Tr14}. Based on the widths and asymmetries of \Ha\ and \Hb\ broad emission line profiles in a sample of type-1 AGNs taken from SDSS DR 16, \citet{Ra22} showed that the BLR gas seems to be virialized. The velocity-resolved lag maps of H$\beta$ broad emission line for Mrk 50 and SBS 1518+593 show characteristic of Keplerian disk or virialized motion \citep{Ba11,Du18a}. Thus, it is likely that the redward shifts of broad emission lines originate from the gravity of the central black hole.

Radiation pressure from accretion disk has significant influences on the stability and dynamics of clouds in BLR \citep[e.g.,][]{Ma08,Ne10,Kr11,Kr12,Na21}. The dynamics of clouds can determine the three-dimensional geometry of BLR \citep{Na21}. However, radiation pressure was not considered in estimating $M_{\rm{RM}}$, and the virial factor had been believed to be only from the geometric effect of BLR. \citet{Lu16} found that the BLR of NGC 5548 could be jointly controlled by the radiation pressure force from accretion disk and the gravity of the central black hole. \citet{Kr11} found that stable orbits of clouds in BLR exist for very sub-Keplerian rotation, for which the radiation pressure force contributes substantially to the force budget. Thus, the radiation pressure force may result in significant influence on the virial factor. Based on redward-shifted \Hb\ and \feii\ broad emission lines for a sample of 1973 $z < 0.8$ SDSS DR5 quasars, \citet{Li22} found a positive correlation of the virial factor with the dimensionless accretion rate or the Eddington ratio. They suggested that the radiation pressure force is a significant contributor to the virial factor, and that the redward shift of \Hb\ broad emission line is mainly from the gravity of the black hole. In this work, more than 8000 SDSS DR7 AGNs with redward-shifted \Hb\ and \Ha\ broad emission lines, out of Table 2 in \citet{Li19}, will be adopted  to investigate the virial factor, relations of the virial factor with other physical quantities, the origin of the redward shifts of the broad Balmer emission lines, and the implications of the $f$ correction.

The structure is as follows. Section~\ref{sec:meth} presents method. Section~\ref{sec:samp} describes sample selection. Section~\ref{sec:anal} presents analysis and results. Section~\ref{sec:quas} is potential influence on quasars at $z\gtrsim 6$. Section~\ref{sec:msig} is potential influence on $M_{\bullet}-\sigma_{\ast}$ map of AGNs. Section~\ref{sec:diss} presents discussion, and Section~\ref{sec:conc} is conclusion. Throughout this paper, we assume a standard cosmology with $H_0=70 \rm{\/\ km \/\ s^{-1} \/\ Mpc^{-1}}$, $\Omega_{\rm{M}}$ = 0.3, and $\Omega_{\rm{\Lambda}}$= 0.7 \citep{Sp07}.

\section{METHOD}\label{sec:meth}
A BLR cloud is subject to gravity of black hole, $F_{\rm{g}}$, and radiation pressure force, $F_{\rm{r}}$, due to central continuum radiation. The total mechanical energy and angular momentum are conserved for the BLR clouds because $F_{\rm{g}}$ and $F_{\rm{r}}$ are central forces. Under various assumptions, $F_{\rm{r}}$ can be calculated for more than hundreds of thousands of lines, with detailed photoionization, radiative transfer, and energy balance calculations \citep[e.g.,][]{Da19}. In principle, $M_{\bullet}$ could be estimated by the BLR cloud motions when the numerical calculation methods give $F_{\rm{r}}$. However, the various assumptions may significantly influence the reliability of $F_{\rm{r}}$. Especially, many unknown physical parameters are likely various for different AGNs. Thus, a new method was proposed to measure $f$ and then $M_{\rm{RM}}$, avoiding to use the averages of the virial factor or the numerical calculation of $F_{\rm{r}}$ \citep{Li17,Li22}.

The virial factor formula in \citet{Li22} was derived from the Schwarzschild metric for clouds in the virialized motion
\begin{equation}
   f = \frac{1}{3} \frac{c^2}{v^2_{\rm{FWHM}}} \left[1-\left(1+z_{\rm{g}} \right)^{-2} \right],
\end{equation}
where the gravitational and transverse Doppler shifts are taken into account. If $v_{\rm{FWHM}}$ is replaced with $\sigma_{\rm{line}}$, $f$ becomes $f_{\sigma}$. As $z_{\rm{g}} \ll 1$ or $r_{\rm{g}}/r_{\rm{BLR}}\ll 1$ for broad emission lines (the gravitational radius $r_{\rm{g}}= GM_{\bullet}/c^2$), we have
\begin{equation}
   f = \frac{2}{3} \frac{c^2}{v^2_{\rm{FWHM}}} z_{\rm{g}}.
\end{equation}

\citet{MJ21} pinpointed that the observed redward shift of the Fe {\sc iii}$\lambda\lambda$ 2039--2113 emission line blend in quasars originates from the gravity of black hole, while these Fe {\sc iii}$\lambda\lambda$ 2039--2113 emission lines are broad emission lines. Furthermore, their redward shifts and line widths follow the gravitational redshift prediction \citep[see Figure 4 in][]{MJ18}. So, the broad emission line position and width should be not only determined by the kinematics of BLR, but also determined by the gravity of black hole. The virialized assumption in measuring $M_{\rm{RM}}$ will ensure that the line position should be governed by the gravity of black hole for the redward shifted broad emission lines in AGNs (this will be tested in the next section). The same method as in \citet{Li17,Li22} was used to estimate the virial factor in \citet{MJ18} (see their Equation 5). Thus, the method in this work, evolved from \citet{Li17}, is reliable, and the assumption of the gravitational redshift is reasonable.

The Schwarzschild metric is valid at the optical BLR scales, and Equation (1) is valid for a disklike BLR \citep[see][]{Li22}. The disklike BLR is preferred by some RM observations of AGNs, e.g., NGC 3516 \citep[e.g.,][]{De10,Fe21a}, and the VLTI instrument GRAVITY observations of quasar 3C 273 \citep{St18}. For rapidly rotating BLR clouds, the relativistic beaming effect can give rise to a profile asymmetry with an enhanced blue side in broad emission lines, i.e., blueshifts of broad emission lines \citep{Me89}. Thus, the relativistic beaming effect should be neglected for the redward shifted broad emission lines, which should be dominated by the gravitational redshift and transverse Doppler effects. The reliability of the redward shift method was confirmed by the consistent masses estimated from Equations (4) and (7) based on 4 broad emission lines for Mrk 110 \citep[see Figure 2 of][]{Li17}. Hereafter, $M_{\rm{RM}}$ denotes $M_{\bullet}$ measured with the RM method and/or the relevant secondary methods, $f_{\rm{g}}$ denotes $\langle f\rangle=1$ for $v_{\rm{FWHM}}$ or $\langle f_{\rm{\sigma}}\rangle=5.5$ for $\sigma_{\rm{line}}$, $M_{\rm{RM}}\equiv M_{\rm{RM}}(f_{\rm{g}}=1)$, the Eddington luminosity $L_{\rm{Edd}}\equiv L_{\rm{Edd}}(f_{\rm{g}}=1)$, the Eddington ratio $L_{\rm{bol}}/L_{\rm{Edd}}\equiv L_{\rm{bol}}/L_{\rm{Edd}}(f_{\rm{g}}=1)$, and $r_{\rm{g}}\equiv r_{\rm{g}}(f_{\rm{g}}=1)$.

\section{SAMPLE SELECTION}\label{sec:samp}
\citet{Li19} reported a comprehensive and uniform sample of 14584 broad-line AGNs with $z < 0.35$ from the SDSS DR7. The stellar continuum was properly removed for each spectrum with significant host absorption line features, and careful analyses of emission line spectra, particularly in the \Ha\ and \Hb\ wavebands, were carried out. The line widths and line centroid wavelengths of the \Ha, \Hb, and \oiii\ spectra are given in Table 2 of \citet{Li19}. The redward shifts of broad emission lines \Hb\ and \Ha\ are defined as
\begin{equation}
\begin{split}
z_{\rm{g}}&=\frac{\lambda_{\rm{b}}\rm{(H)}}{\lambda_{\rm{n}}\rm{(H)}}-1 \\
&=\frac{{\lambda_{\rm{b}}\rm{(H)}}}{\lambda_{\rm{n}}\rm{(\oiii)}}
\frac{\lambda_{\rm{0}}\rm{(\oiii)}}{\lambda_{\rm{0}}\rm{(H)}}-1,
\end{split}
\end{equation}
where ${\lambda_{\rm{b}}}$ is the centroid wavelength of broad emission line corrected by the cosmological redshift $z_{\rm{SDSS}}$ given by the SDSS site \citep{Li19}, ${\lambda_{\rm{n}}}$ is the centroid wavelength of narrow emission line corrected by $z_{\rm{SDSS}}$, and $\lambda_{\rm{0}}$ is the vacuum wavelength of spectrum line ($\lambda_{\rm{0}} =4862.68$ $\rm{\AA}$ for \Hb, $\lambda_{\rm{0}} =6564.61$ $\rm{\AA}$ for \Ha, and $\lambda_{\rm{0}}=5008.24$ $\rm{\AA}$ for \oiii $\lambda 5007$)\footnote{https://classic.sdss.org/dr6/algorithms/linestable.html}.

Because of the absence of the uncertainty of ${\lambda_{\rm{n}}}$ for \Hb\ in Table 2 of \citet{Li19}, and in order to unify standard of estimating $z_{\rm{g}}$ for the broad \Hb\ and \Ha, the \oiii $\lambda 5007$ line is used in Equation (3). First, one of choice criteria is AGN's flag = 0, which means no emission line with multiple peaks \citep{Li19}, because that the multiple peaks of emisson lines may be from dual AGNs \citep[e.g.,][]{Wa09}. Second, AGNs are selected on the basis of $z_{\rm{g}} > 0$ and $z_{\rm{g}} - \sigma(z_{\rm{g}})> 0$ for the broad \Hb\ and \Ha, where $\sigma(z_{\rm{g}})$ is the uncertainty of $z_{\rm{g}}$. Third, AGNs are selected on the basis of $v_{\rm{FWHM}} > 0$ and $v_{\rm{FWHM}} - \sigma(v_{\rm{FWHM}})> 0$ for the broad \Hb\ and \Ha, where $\sigma(v_{\rm{FWHM}})$ is the uncertainty of $v_{\rm{FWHM}}$. The selection conditions of $z_{\rm{g}} > 0$ and $z_{\rm{g}} - \sigma(z_{\rm{g}})> 0$ make sure that the shifts of broad emission lines are redward within $1\sigma$ uncertainties. Because the empirical $r_{\rm{BLR}}$--$L_{\rm{5100}}$ relation is established for broad emission line \Hb, the relevant researches on the virial factor are made with the broad \Hb\ and \Ha\ in this work. 9185 AGNs are selected out of the 14584 AGNs as Sample 1 only for the broad \Hb. 9271 AGNs are selected out of the 14584 AGNs as Sample 2 only for the broad \Ha. The cross-identified AGNs in Samples 1 and 2 are used as Sample 3 that contains 8169 AGNs with $z_{\rm{g}}$ of the broad \Hb\ and \Ha.

Some physical quantities are taken or estimated from Table 2 in \citet{Li19}, including $v_{\rm{FWHM}}$(\Hb), $v_{\rm{FWHM}}$(\Ha), $z_{\rm{g}}$(\Hb), $z_{\rm{g}}$(\Ha), $L_{\rm{5100}}$, $M_{\rm{RM}}$, $L_{\rm{bol}}/L_{\rm{Edd}}$, and the dimensionless accretion rate $\mathscr{\dot M}_{f_{\rm{g}}=1}$. The bolometric luminosity $L_{\rm{bol}}$ was estimated in \citet{Li19} using $L_{\rm{bol}}=9.8 L_{\rm{5100}}$ \citep{Mc04}. The details of samples are listed in Tables~\ref{sam1}--\ref{sam3}. The virial factors, $f$(\Hb) and $f$(\Ha), are estimated by Equation (1) for the broad \Hb\ and \Ha\ (see Tables~\ref{sam1}--\ref{sam3}). $\mathscr{\dot M}_{f_{\rm{g}}=1} = L_{\rm{bol}}/L_{\rm{Edd}}/\eta$, where $\eta$ is the efficiency of converting rest-mass energy to radiation. Hereafter, in addition to special statement, we adopt $\eta =0.038$ \citep{Du15}.

For our selected AGNs, $L_{\rm{5100}}$ spans four orders of magnitude, $M_{\rm{RM}}$ spans more than four orders of magnitude, and $L_{\rm{bol}}/L_{\rm{Edd}}$ spans more than three orders of magnitude. These parameters cover at least one order of magnitude wider than those in \citet{Li22}. The measured values of $f$ span more than three orders of magnitude, which cover at least one order of magnitude wider than those in \citet{Li22}. These much wider parameters can ensure that this work is feasible.

\begin{landscape}
\begin{deluxetable}{ccccccccccccc}
  \tablecolumns{13}
  \setlength{\tabcolsep}{1.6pt}
  \tablewidth{0pc}
  \tablecaption{The Relevant Parameters for 9185 AGNs in SDSS DR7 for Sample 1}
  \tabletypesize{\scriptsize}
  \tablehead{\colhead{Designation} & \colhead{$\frac{v_{\rm{FWHM}}(\rm{H\beta})}{\rm{km \/\ s^{-1}}}$}  & \colhead{$z_{\rm{g}}$(\Hb)} & \colhead{$\log L_{\rm{5100}}$} & \colhead{$\log \frac{M_{\rm{RM}}}{M_{\odot}}$} & \colhead{$ \log \frac{L_{\rm{bol}}}{L_{\rm{Edd}}}$} & \colhead{$f$(\Hb)}  & \colhead{$\log \mathscr{\dot M}_{f_{\rm{g}}=1}$}  & \colhead{$\frac{r_{\rm{BLR}}}{r_{\rm{g}}}$}&  \colhead{$R_{\rm{FeII}}$} & \colhead{$\log \frac{M_{\rm{RM}}}{M_{\odot}}^{\dag}$}  & \colhead{$\frac{r_{\rm{BLR}}}{r_{\rm{g}}}^{\dag}$} & \colhead{$\log \mathscr{\dot M}_{f_{\rm{g}}=1}^{\dag}$} \\
  \colhead{(1)}  &\colhead{(2)}  &\colhead{(3)}  &\colhead{(4)}  &\colhead{(5)}  &\colhead{(6)}  &  \colhead{(7)}  &\colhead{(8)} &\colhead{(9)} &\colhead{(10)} &\colhead{(11)} &\colhead{(12)} &\colhead{(13)} }

  \startdata
J000048.16-095404.0 & 2132.9$\pm$71.0 & 0.00089$\pm$0.00005 & 43.39$\pm$0.00 & 7.26 & -1.086 & 11.7$\pm$0.8 & 0.334 & 15381.3 & -999 & -999&-999 & -999 \\
J000102.19-102326.9 & 4695.0$\pm$69.2 & 0.00062$\pm$0.00049 & 44.36$\pm$0.00 & 8.46 & -1.356 & 1.7$\pm$0.0 & 0.064 & 3191.6 &0.188 & 8.49& 3191.7&0.037 \\
J000154.29+000732.5 & 1813.8$\pm$98.9 & 0.00046$\pm$0.00023 & 43.48$\pm$0.00 & 7.14 & -0.972 & 8.4$\pm$0.9 & 0.448 & 22644.5 &0.115 &7.27 &22645.3 &0.322 \\
... & ...  & ... &  ... & ... &  ... & ... & ... & ... & ... & ... & ... & ...\\

\enddata
\tablecomments{Column 1: object name; Column 2: $v_{\rm{FWHM}}$ of \Hb\ broad emission line; Column 3: the redward shift of \Hb; Column 4: logarithm of $L_{\rm{5100}}$ in units of $\rm{erg \/\ s^{-1}}$; Column 5: logarithm of $M_{\rm{RM}}$ in units of $M_{\odot}$; Column 6: logarithm of $L_{\rm{bol}}/L_{\rm{Edd}}$; Column 7: $f$ estimated from $v_{\rm{FWHM}}$ of \Hb; Column 8: logarithm of $\mathscr{\dot M}_{f_{\rm{g}}=1}$; Column 9: $r_{\rm{BLR}}$ in units of $r_{\rm{g}}$, where $r_{\rm{BLR}}=33.65L^{0.533}_{44}$ light-days with $L_{44}=L_{\rm{5100}}/(10^{44}\/\ \rm{erg \/\ s^{-1}})$. \textbf{Column 10: $R_{\rm{FeII}}$ is the line ratio of \feii\ to \Hb.} Columns 2--6 and 10 are taken from Table 2 of \citet{Li19} or converted from the relevant quantities in Table 2 of \citet{Li19}. -999 denotes no data of $R$(\feii), which results in no data for the latter three quantities. $\dag$ denotes the values estimated from Equations (6) and (7).  \\(This table is available in its entirety in machine-readable form.)}
\label{sam1}
\end{deluxetable}
\end{landscape}

\begin{landscape}
\begin{deluxetable}{ccccccccccc}
  \tablecolumns{11}
  \setlength{\tabcolsep}{1.6pt}
  \tablewidth{0pc}
  \tablecaption{The Relevant Parameters for 9271 AGNs in SDSS DR7 for Sample 2}
  \tabletypesize{\scriptsize}
  \tablehead{\colhead{Designation} & \colhead{$\frac{v_{\rm{FWHM}}(\rm{H\alpha})}{\rm{km \/\ s^{-1}}}$} & \colhead{$z_{\rm{g}}$(\Ha)}& \colhead{$z_{\rm{g}}$(\Ha)(b-n)} &
\colhead{$\log L_{\rm{5100}}$} & \colhead{$\log \frac{M_{\rm{RM}}}{M_{\odot}}$} & \colhead{$\log \frac{L_{\rm{bol}}}{L_{\rm{Edd}}}$} & \colhead{$f$(\Ha)} & \colhead{$f$(\Ha)(b-n)} & \colhead{$\log \mathscr{\dot M}_{f_{\rm{g}}=1}$} & \colhead{$\frac{r_{\rm{BLR}}}{r_{\rm{g}}}$} \\
  \colhead{(1)}  &\colhead{(2)}  &\colhead{(3)}  &\colhead{(4)}  &\colhead{(5)}  &\colhead{(6)}&
  \colhead{(7)}  &\colhead{(8)}  &\colhead{(9)}  &\colhead{(10)} &\colhead{(11)} }

  \startdata
J000048.16-095404.0 & 2132.9$\pm$282.9 & 0.00089$\pm$0.00022 & 0.00063$\pm$0.00022 & 43.39$\pm$0.00 & 7.26 & -1.086 & 11.7$\pm$3.1 & 8.3$\pm$2.2 & 0.334 & 15381.3 \\
J000102.19-102326.9 & 4695.0$\pm$172.5 & 0.00062$\pm$0.00050 & 0.00073$\pm$0.00050 & 44.36$\pm$0.00 & 8.46 & -1.356 & 1.7$\pm$0.1 & 2.0$\pm$0.1 & 0.064 & 3191.6 \\
J000111.15-100155.5 & 1937.4$\pm$84.2 & 0.00015$\pm$0.00012 & 0.00013$\pm$0.00012 &
43.15$\pm$0.04 & 6.37 & -0.327 & 2.4$\pm$0.2 & 2.1$\pm$0.2 & 1.093 & 88935.0 \\
... & ... & ... & ... & ... &  ... & ... & ... & ... & ... &  ... \\

\enddata
\tablecomments{Column 1: object name; Column 2: $v_{\rm{FWHM}}$ of \Ha\ broad emission line; Column 3: the redward shift of broad \Ha\ with respect to \oiii$\lambda$5007; Column 4: the redward shift of broad \Ha\ with respect to narrow \Ha; Column 5: logarithm of $L_{\rm{5100}}$
in units of $\rm{erg \/\ s^{-1}}$; Column 6: logarithm of $M_{\rm{RM}}$ in units of $M_{\odot}$; Column 7: logarithm of $L_{\rm{bol}}/L_{\rm{Edd}}$; Column 8: the virial factor estimated from $v_{\rm{FWHM}}$ of broad \Ha\ and $z_{\rm{g}}$(\Ha); Column 9: the virial factor estimated from $v_{\rm{FWHM}}$ of broad \Ha\ and $z_{\rm{g}}$(\Ha)(b-n); Column 10: logarithm of $\mathscr{\dot M}_{f_{\rm{g}}=1}$; Column 11: $r_{\rm{BLR}}$ in units of $r_{\rm{g}}$, where $r_{\rm{BLR}}=33.65L^{0.533}_{44}$ light-days with $L_{44}=L_{\rm{5100}}/(10^{44}\/\ \rm{erg \/\ s^{-1}})$. Columns 2--7 are taken from Table 2 of \citet{Li19} or converted from the relevant quantities in Table 2 of \citet{Li19}. \\(This table is available in its entirety in machine-readable form.)}
\label{sam2}
\end{deluxetable}
\end{landscape}

\begin{landscape}
\begin{deluxetable}{cccccccccccccccc}
  \tablecolumns{16}
  \setlength{\tabcolsep}{1.6pt}
  \tablewidth{0pc}
  \tablecaption{The Relevant Parameters for 8169 AGNs in SDSS DR7 for Sample 3}
  \tabletypesize{\scriptsize}
  \tablehead{\colhead{Designation}  & \colhead{$\frac{v_{\rm{FWHM}}(\rm{H\beta})}{\rm{km \/\ s^{-1}}}$}  & \colhead{$z_{\rm{g}}$(\Hb)} & \colhead{$\frac{v_{\rm{FWHM}}(\rm{H\alpha})}{\rm{km \/\ s^{-1}}}$}  & \colhead{$z_{\rm{g}}$(\Ha)} & \colhead{$\log L_{\rm{5100}}$} & \colhead{$\log \frac{M_{\rm{RM}}}{M_{\odot}}$} & \colhead{$\log \frac{L_{\rm{bol}}}{L_{\rm{Edd}}}$} & \colhead{$f$(\Hb)}  & \colhead{$f$(\Ha)} & \colhead{$\log \mathscr{\dot M}_{f_{\rm{g}}=1}$}  & \colhead{$\frac{r_{\rm{BLR}}}{r_{\rm{g}}}$} & \colhead{$R_{\rm{FeII}}$} & \colhead{$\log \frac{M_{\rm{RM}}}{M_{\odot}}^{\dag}$}  & \colhead{$\frac{r_{\rm{BLR}}}{r_{\rm{g}}}^{\dag}$} & \colhead{$\log \mathscr{\dot M}_{f_{\rm{g}}=1}^{\dag}$} \\
  \colhead{(1)}  &\colhead{(2)}  &\colhead{(3)}  &\colhead{(4)}  &\colhead{(5)}  &\colhead{(6)}  &
  \colhead{(7)}  &\colhead{(8)}  &\colhead{(9)}  &\colhead{(10)}&
  \colhead{(11)}  &\colhead{(12)} &\colhead{(13)}  &\colhead{(14)}&
  \colhead{(15)}  &\colhead{(16)} }

  \startdata
J000048.16-095404.0 & 2132.9$\pm$71.0 & 0.00089$\pm$0.00005 & 2132.9$\pm$282.9 & 0.00089$\pm$0.00022 & 43.39$\pm$0.00 & 7.26 & -1.086 & 11.7$\pm$0.8 & 11.7$\pm$3.1 & 0.334 & 15381.3 & -999 & -999&-999 & -999\\
J000102.19-102326.9 & 4695.0$\pm$69.2 & 0.00062$\pm$0.00049 & 4695.0$\pm$172.5 & 0.00062$\pm$0.00050 & 44.36$\pm$0.00 & 8.46 & -1.356 & 1.7$\pm$0.0 & 1.7$\pm$0.1 & 0.064 & 3191.6 & 0.188 &8.49& 6663.8& 0.037 \\
J000154.29+000732.5 & 1813.8$\pm$98.9 & 0.00046$\pm$0.00023 & 1813.8$\pm$178.9 & 0.00046$\pm$0.00023 & 43.48$\pm$0.00 & 7.14 & -0.972 & 8.4$\pm$0.9 & 8.4$\pm$1.7 & 0.448 & 22644.5 & 0.115 & 7.27 &52991.6& 0.322 \\
... & ... & ... & ... &  ... & ... &  ... & ... & ... & ... &  ... & ... & ... & ... &  ... & ... \\

\enddata
\tablecomments{Column 1: object name; Column 2: $v_{\rm{FWHM}}$ of \Hb\ broad emission line; Column 3: the redward shift of \Hb; Column 4: $v_{\rm{FWHM}}$ of \Ha\ broad emission line; Column 5: the redward shift of \Ha; Column 6: logarithm of $L_{\rm{5100}}$ in units of $\rm{erg \/\ s^{-1}}$; Column 7: logarithm of $M_{\rm{RM}}$ in units of $M_{\odot}$; Column 8: logarithm of $L_{\rm{bol}}/L_{\rm{Edd}}$; Column 9: $f$ estimated from $v_{\rm{FWHM}}$ of \Hb; Column 10: $f$ estimated from $v_{\rm{FWHM}}$ of \Ha; Column 11: logarithm of $\mathscr{\dot M}_{f_{\rm{g}}=1}$; Column 12: $r_{\rm{BLR}}$ in units of $r_{\rm{g}}$, where $r_{\rm{BLR}}=33.65L^{0.533}_{44}$ light-days with $L_{44}=L_{\rm{5100}}/(10^{44}\/\ \rm{erg \/\ s^{-1}})$. \textbf{Column 13: $R_{\rm{FeII}}$ is the line ratio of \feii\ to \Hb.} Columns 2--8 and 13 are taken from Table 2 of \citet{Li19} or converted from the relevant quantities in Table 2 of \citet{Li19}. -999 and $\dag$ are same as in Table~\ref{sam1}. \\(This table is available in its entirety in machine-readable form.)}
\label{sam3}
\end{deluxetable}
\end{landscape}

\section{ANALYSIS AND RESULTS}\label{sec:anal}
In order to \textbf{study the correlation between $f$, $\mathscr{\dot M}_{f_{\rm{g}}=1}$, $L_{\rm{5100}}$, and $v_{\rm{FWHM}}$, as well as $z_{\rm{g}}$ and $r_{\rm{BLR}}/r_{\rm{g}}$, we will perform the Spearman's rank test and/or the Pearson's correlation analysis. The bisector linear regression \citep{Is90} is performed} to obtain the slope and intercept coefficients of $y=a+bx$ \textbf{in fitting our samples, if needed for some quantities. The} partial correlation analysis is \textbf{used} to further verify the presence of \textbf{correlation between $f$ and $\mathscr{\dot M}_{f_{\rm{g}}=1}$}. All correlation analyses are calculated in log-space. The SPEAR \citep{Pr92} is used to calculate the Spearman's rank correlation coefficient $r_{\rm{s}}$ and the $p$-value $P_{\rm{s}}$ of the hypothesis test. The PEARSN \citep{Pr92} is used to give the Pearson's correlation coefficient $r$ and the $p$-value $P$ of the hypothesis test.

\begin{figure}
  \begin{center}
    \includegraphics[angle=0,scale=0.26]{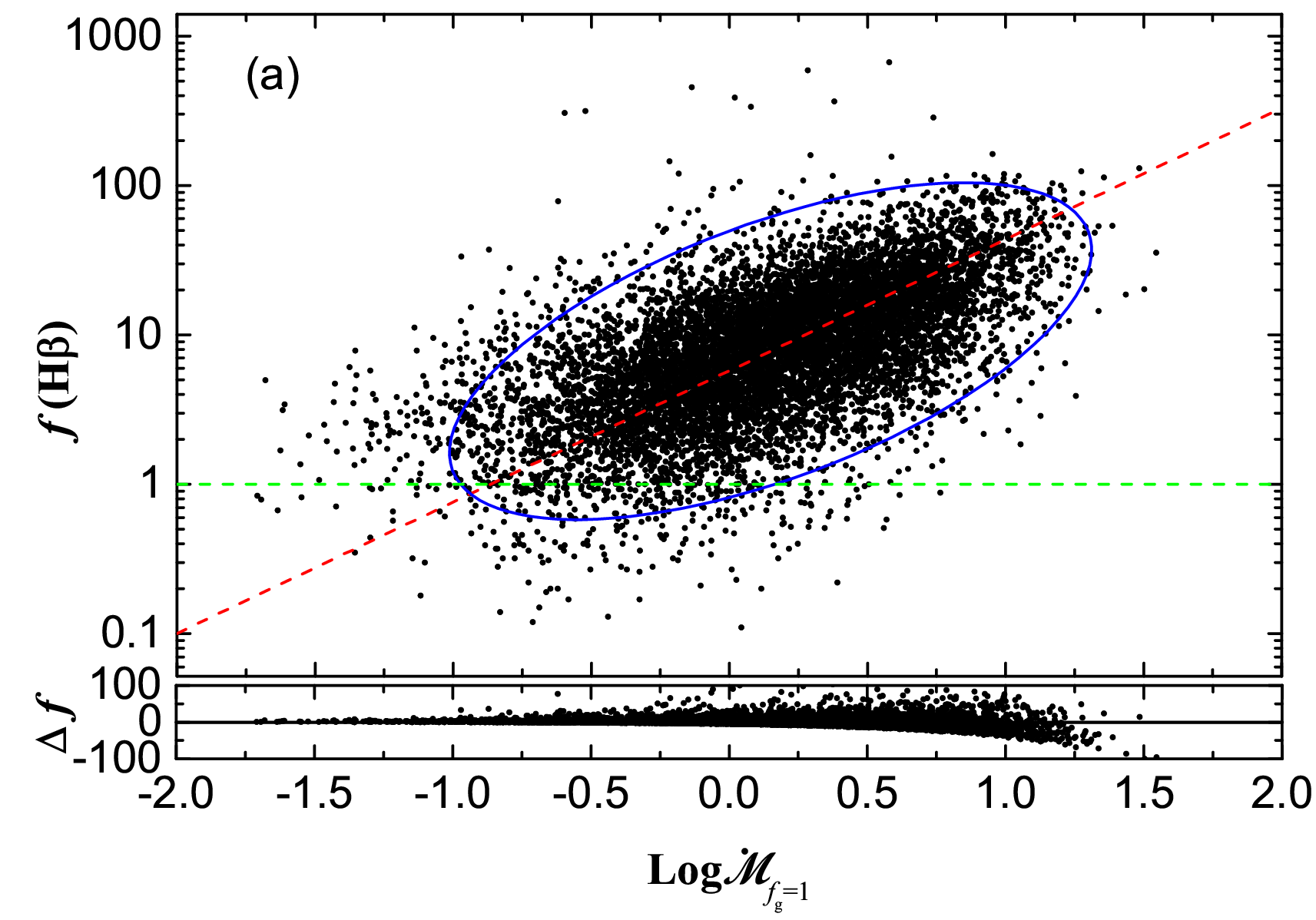}
    \includegraphics[angle=0,scale=0.26]{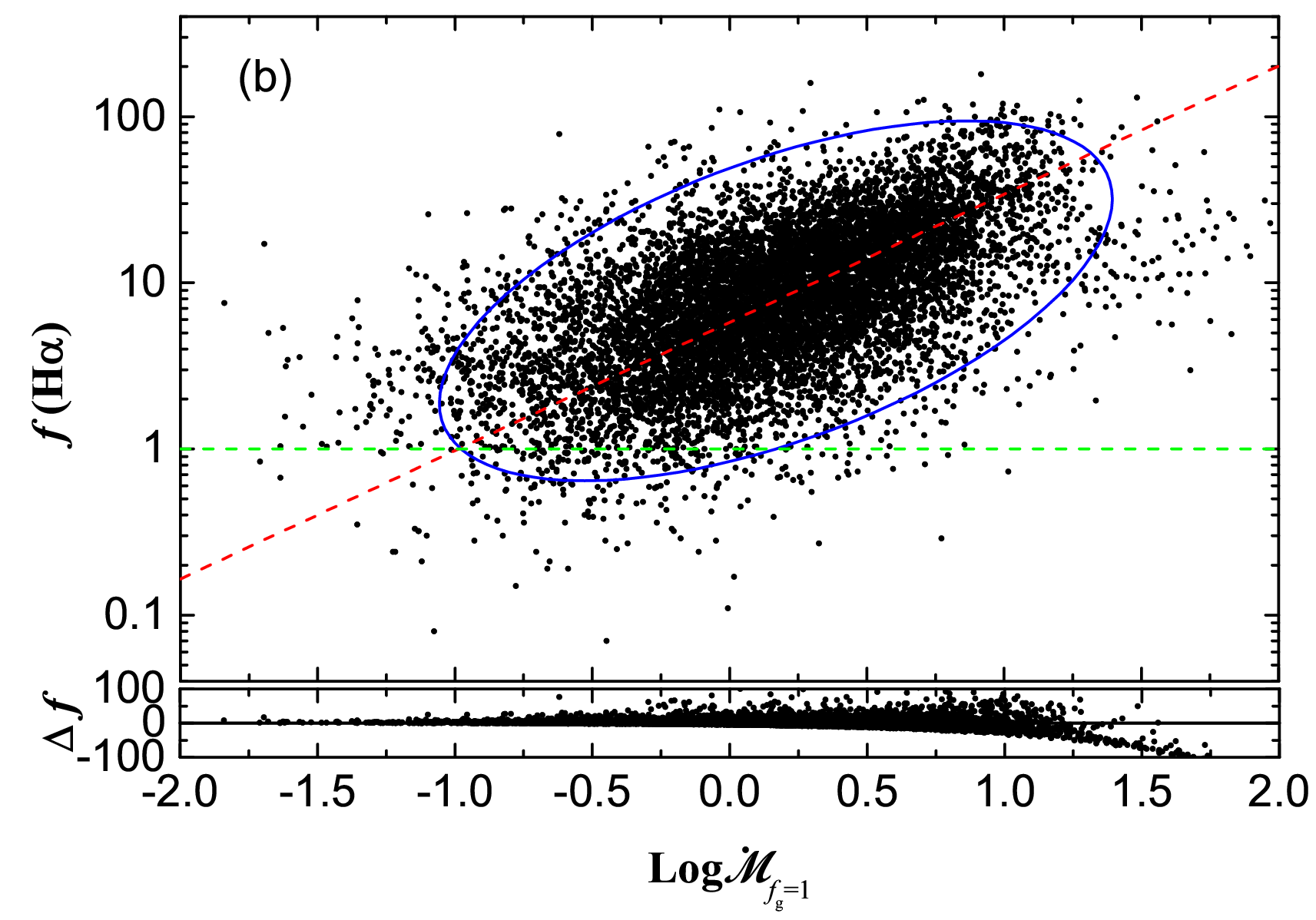}
    \includegraphics[angle=0,scale=0.26]{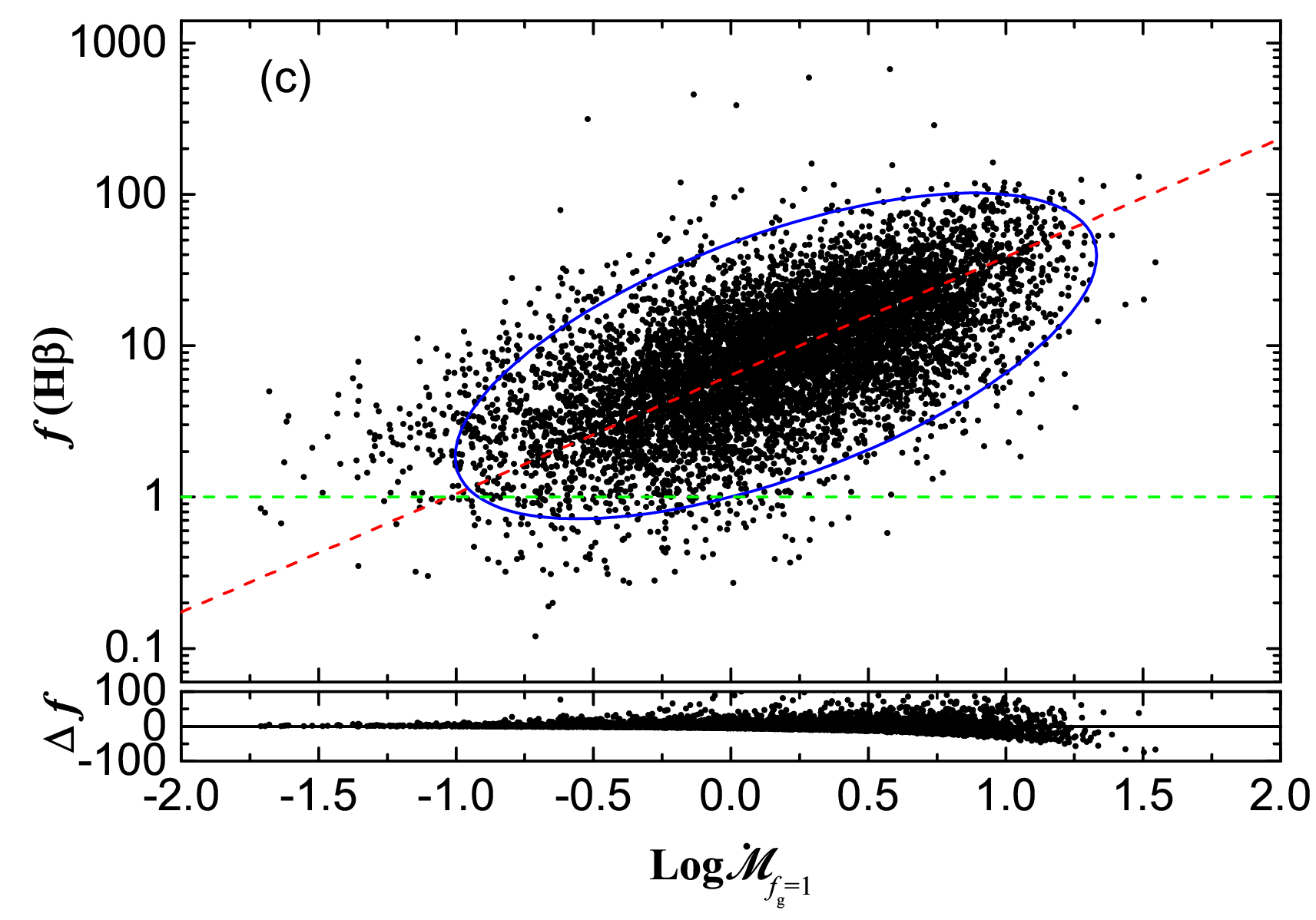}
    \includegraphics[angle=0,scale=0.26]{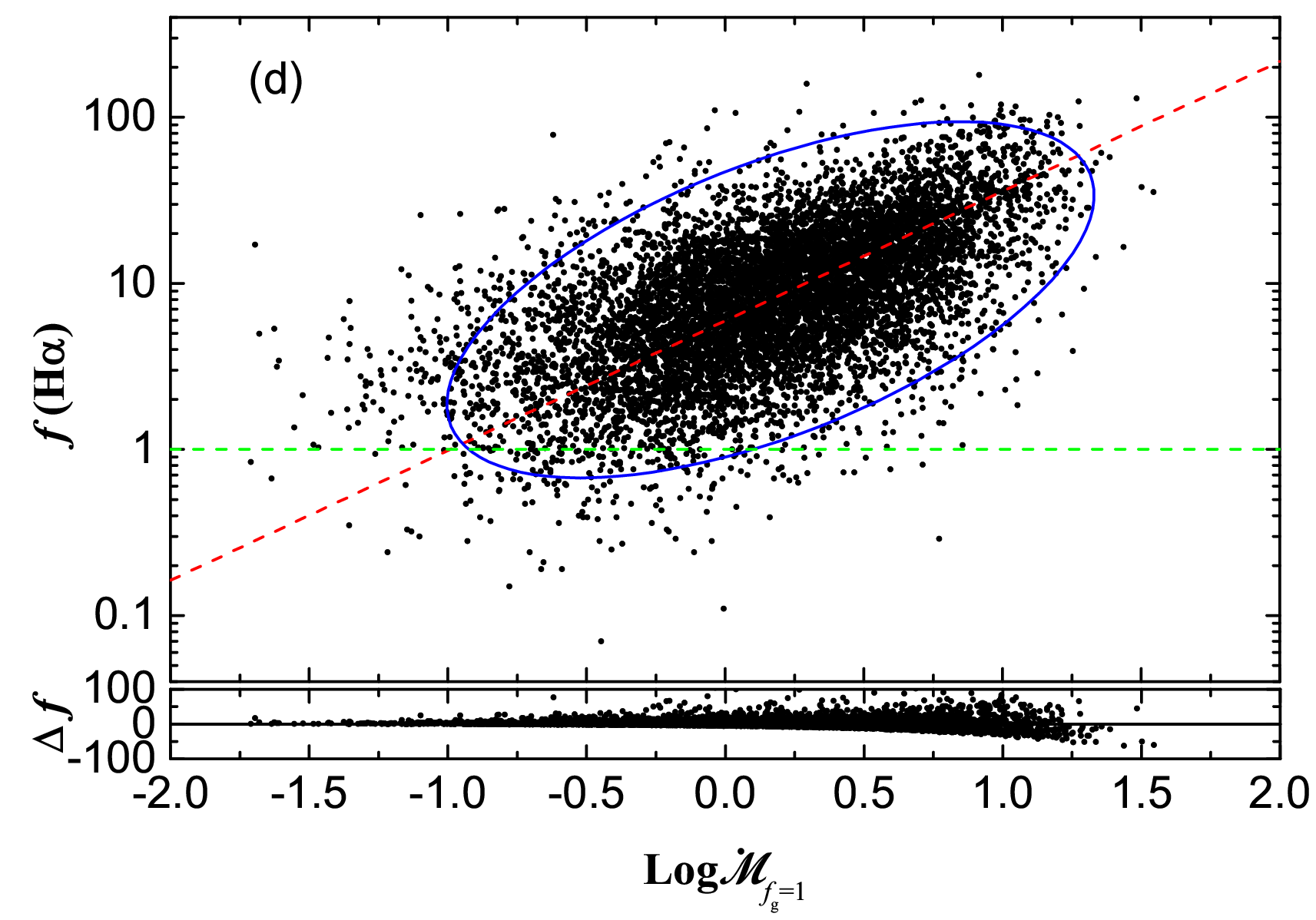}
  \end{center}
  \caption{$f$ vs. $\mathscr{\dot M}_{f_{\rm{g}}=1}$. Panel ($a$): for \Hb\ of 9185 AGNs in Sample 1. Panel ($b$): for \Ha\ of 9271 AGNs in Sample 2. Panel ($c$): for \Hb\ of 8169 AGNs in Sample 3. Panel ($d$): for \Ha\ of 8169 AGNs in Sample 3. The dashed green line denotes $f_{\rm{g}}=1$ for $v_{\rm{FWHM}}$. The dashed red line denotes the best bisector linear fit. The blue solid line denotes the 95\% confidence ellipse. $\Delta f$ is the fitting residuals.}
  \label{fig:fac-rat}
\end{figure}

The Spearman's rank correlation test is run for Samples 1--3, and the analysis results are listed in Table~\ref{spearman}. There are positive correlations between the virial factor and $\mathscr{\dot M}_{f_{\rm{g}}=1}$ or $L_{\rm{bol}}/L_{\rm{Edd}}$ for Samples 1--3 (see Figure~\ref{fig:fac-rat} and Table~\ref{spearman}). The results from the Pearson's correlation analysis are listed in Table~\ref{pearson}. The bisector regression fit can give $a$ and $b$, as well \textbf{as} their uncertainties $\Delta a$ and $\Delta b$, but it does not take into account the obervational errors of data \citep{Is90}. So, based on Monte Carlo simulated data sets from the obervational values and errors, we calculate the best parameters using the bisector regression, and repeat this procedure $10^4$ times to generate the distributions of $a_{\rm{MC}}$, $b_{\rm{MC}}$, $\Delta a_{\rm{MC}}$, and $\Delta b_{\rm{MC}}$. The means of the $a_{\rm{MC}}$ and $b_{\rm{MC}}$ distributions are taken to be the final best parameters of $a$ and $b$, respectively. The corresponding uncertainties are given by the combinations of the means of the $\Delta a_{\rm{MC}}$ and $\Delta b_{\rm{MC}}$ distributions with the standard deviations of the $a_{\rm{MC}}$ and $b_{\rm{MC}}$ distributions, respectively. The bisector regression is run for $\log f=a+b \log \mathscr{\dot M}_{f_{\rm{g}}=1}$, and the fitting results are
\begin{subequations}
\begin{align}
  \log f &= 0.76(\pm0.01)+0.88(\pm0.01) \log \mathscr{\dot M}_{f_{\rm{g}}=1}, \label{Za} \\
  \log f &= 0.76(\pm0.01)+0.77(\pm0.01) \log \mathscr{\dot M}_{f_{\rm{g}}=1}, \label{Zb} \\
  \log f &= 0.80(\pm0.01)+0.78(\pm0.01) \log \mathscr{\dot M}_{f_{\rm{g}}=1}, \label{Zc} \\
  \log f &= 0.77(\pm0.01)+0.78(\pm0.01) \log \mathscr{\dot M}_{f_{\rm{g}}=1}, \label{Zd}
\end{align}
\end{subequations}
where the $p$-values of the hypothesis test are $<10^{-40}$, and in the fittings, the uncertainty of $\log \mathscr{\dot M}_{f_{\rm{g}}=1}$ is taken to be 0.4 determined by the uncertainty of 0.4 dex usually used in $M_{\rm{RM}}$. Equations (4$a$)--(4$d$) correspond to the best fits to data sets ($f$,$\mathscr{\dot M}_{f_{\rm{g}}=1}$) for \Hb\ in Sample 1, \Ha\ in Sample 2, \Hb\ in Sample 3, and \Ha\ in Sample 3, respectively. It is clear that $f\propto \mathscr{\dot M}_{f_{\rm{g}}=1}^{0.8-0.9}$, and then $f\propto (L_{\rm{bol}}/L_{\rm{Edd}})^{0.8-0.9}$.

Since $f$ may be affected by $F_{\rm{r}}$, it is possible that $f$ is correlated with $L_{\rm{5100}}$. In fact, there are weak correlations between $f$ and $L_{\rm{5100}}$ (see Tables~\ref{spearman}--\ref{pearson}). Also, the dependence of $f$ and $\mathscr{\dot M}_{f_{\rm{g}}=1}$ on $v_{\rm{FWHM}}$ may result in a false correlation. Thus, the partial correlation analysis is needed to test the $\log f$--$\log \mathscr{\dot M}_{f_{\rm{g}}=1}$ correlation when excluding the influence of $v_{\rm{FWHM}}$ and/or $L_{\rm{5100}}$. Based on the Pearson's correlation coefficients in Table~\ref{pearson}, the 1st order partial correlation analysis gives a confidence level of $>$ 99.99\% for the $\log f$--$\log \mathscr{\dot M}_{f_{\rm{g}}=1}$ correlation when excluding the dependence on $L_{\rm{5100}}$ or $v_{\rm{FWHM}}$ (see Table~\ref{partial}). The 2nd order partial correlation analysis gives a confidence level of $>$ 99.99\% for the $\log f$--$\log \mathscr{\dot M}_{f_{\rm{g}}=1}$ correlation when excluding the dependence on $v_{\rm{FWHM}}$ and $L_{\rm{5100}}$, except for the broad \Ha\ in Sample 3 at a confidence level of 99.84\% (see Table~\ref{partial}). Thus, the positive correlation exists between $f$ and $\mathscr{\dot M}_{f_{\rm{g}}=1}$. This positive correlation is qualitatively consistent with the logical expectation when the overall effect of $F_{\rm{r}}$ on the BLR clouds is taken into account to estimate $M_{\rm{RM}}$. In addition, $f>f_{\rm{g}}=1$ for \Hb\ and \Ha\ in most of AGNs (see Figure~\ref{fig:fac-rat}): $>$ 96.5\% for \Hb\ in Sample 1, $>$ 97.2\% for \Ha\ in Sample 2, and $>$ 97.7\% for \Hb\ and $>$ 97.4\% for \Ha\ in Sample 3.

\begin{figure}
  \begin{center}
    \includegraphics[angle=0,scale=0.26]{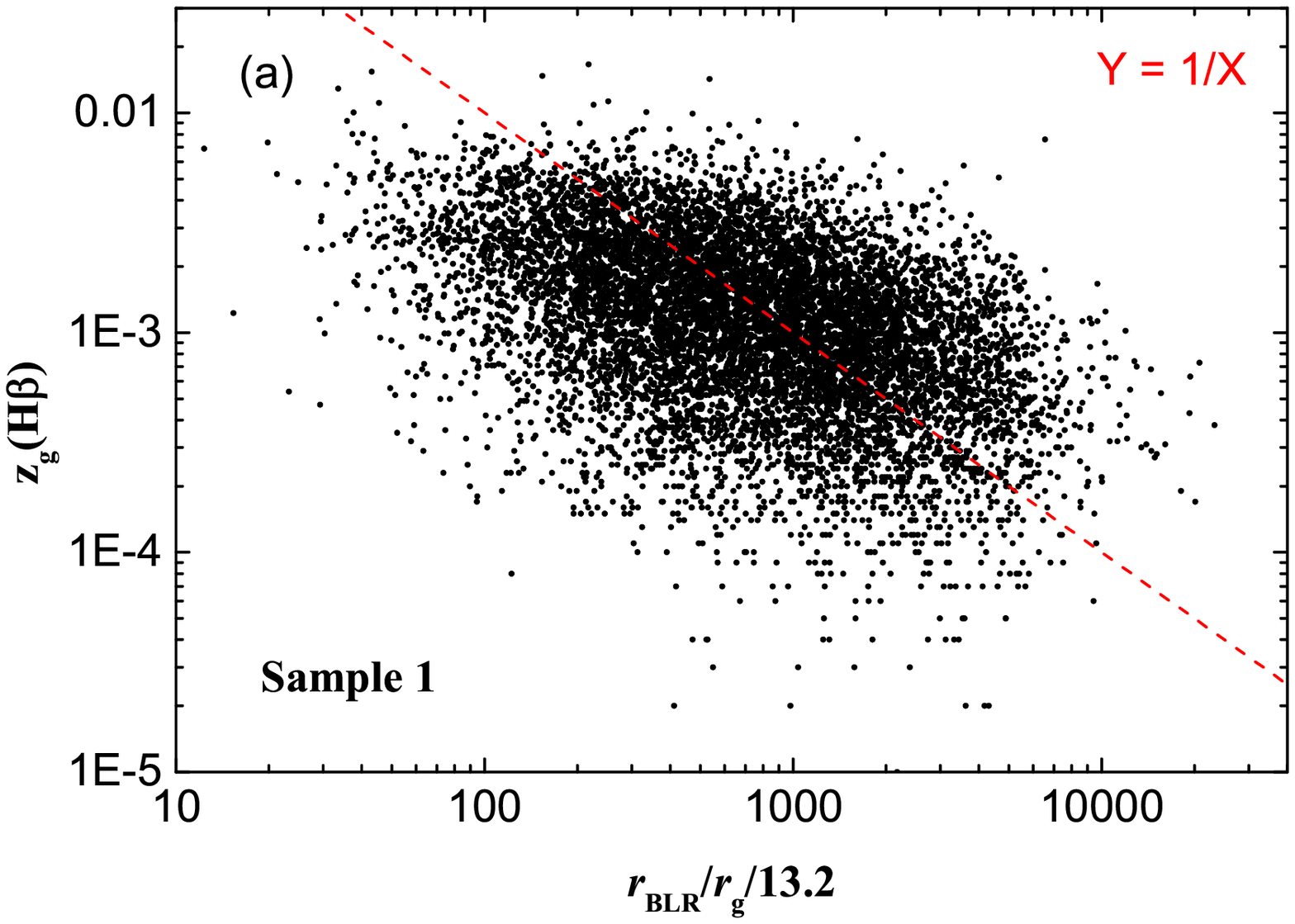}
    \includegraphics[angle=0,scale=0.26]{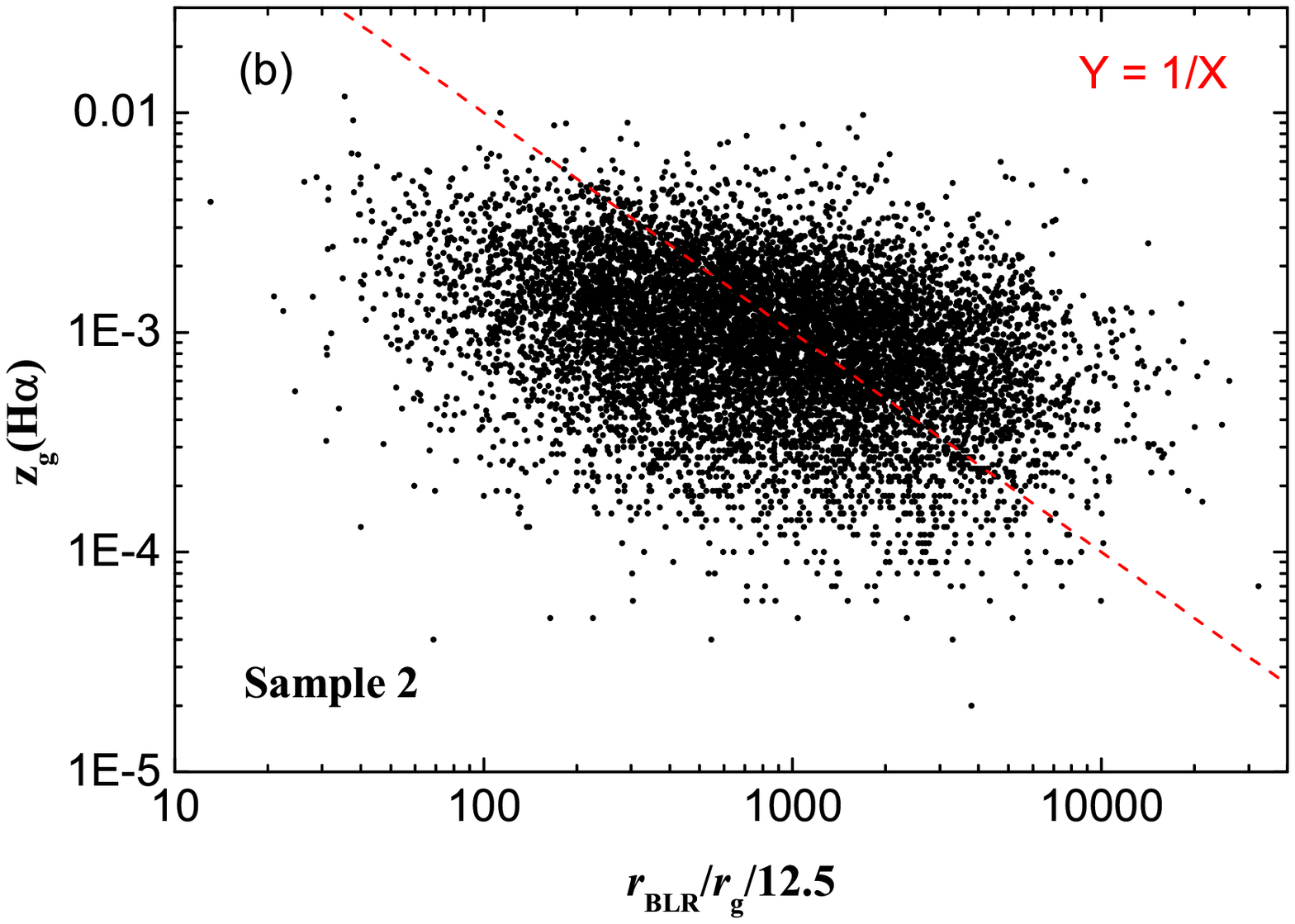}
    \includegraphics[angle=0,scale=0.26]{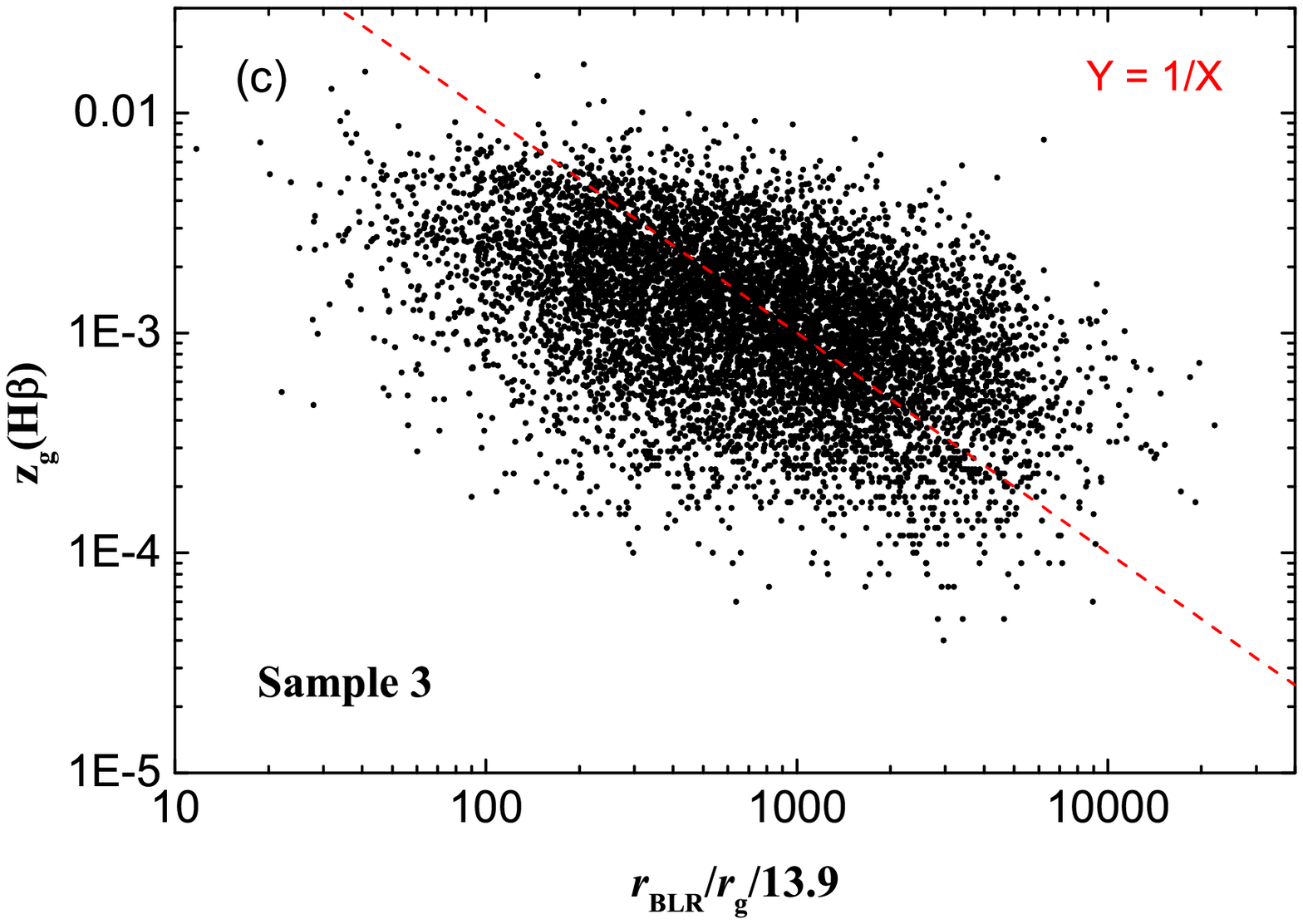}
    \includegraphics[angle=0,scale=0.26]{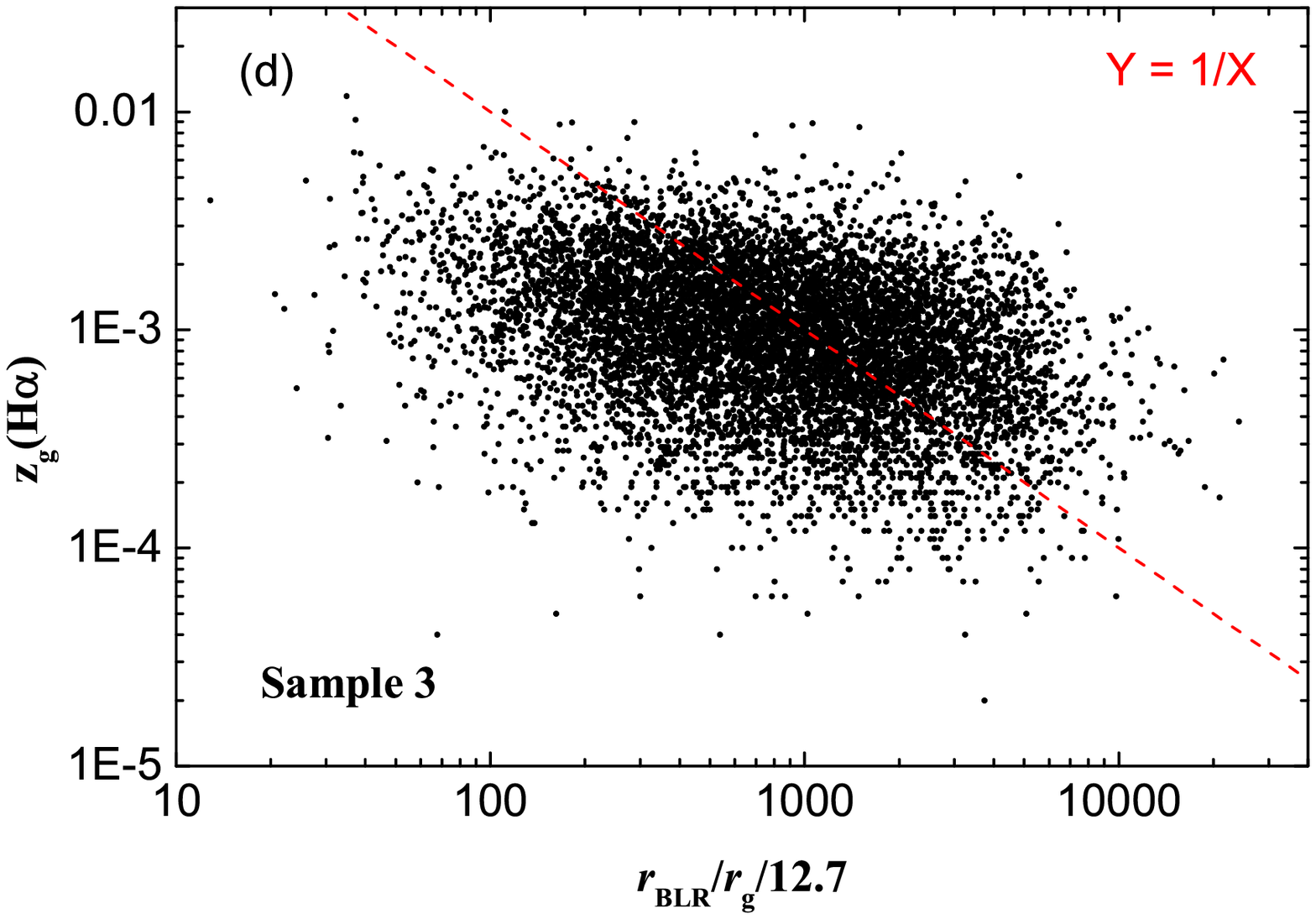}
  \end{center}
  \caption{Panel ($a$): \Hb\ shift $z_{\rm{g}}$ vs. $r_{\rm{BLR}}/r_{\rm{g}}$ corrected by $\langle f\rangle=13.2$ for Sample 1. Panel ($b$): \Ha\ shift $z_{\rm{g}}$ vs. $r_{\rm{BLR}}/r_{\rm{g}}$ corrected by $\langle f\rangle=12.5$ for Sample 2. Panel ($c$): \Hb\ shift $z_{\rm{g}}$ vs. $r_{\rm{BLR}}/r_{\rm{g}}$ corrected by $\langle f\rangle=13.9$ for Sample 3. Panel ($d$): \Ha\ shift $z_{\rm{g}}$ vs. $r_{\rm{BLR}}/r_{\rm{g}}$ corrected by $\langle f\rangle=12.7$ for Sample 3. The Spearman test shows negative correlations between these two physical quantities.}
  \label{fig:shif-rad}
\end{figure}

In order to test the gravitational origin of the redward shift of broad emission line, we
compare $z_{\rm{g}}$ to $r_{\rm{BLR}}/r_{\rm{g}}$, the dimensionless radius of BLR in units of $r_{\rm{g}}$. The Spearman's rank correlation test shows negative correlation between $z_{\rm{g}}$ and $r_{\rm{BLR}}/r_{\rm{g}}$ (see Table~\ref{spearman}). This negative correlation is qualitatively consistent with the expectation that $z_{\rm{g}}$ is mainly from the gravity of the central black hole, because that $M_{\rm{RM}}$ can not be corrected individually for each AGN due to the absence of the individual virial factor that is independent of $z_{\rm{g}}$. So, we can make the overall correction of $M_{\rm{RM}}$ to be $M_{\rm{RM}}/\langle f\rangle$, where $\langle f\rangle$ is the average of $f$ in Samples 1--3 (see Figure~\ref{fig:shif-rad}). This overall correction is equivalent to the parallel shift of the data point in Figure~\ref{fig:shif-rad}. The negative correlation expectation is basically consistent with the trend between $z_{\rm{g}}$ and $r_{\rm{BLR}}/r_{\rm{g}}/\langle f\rangle$ in Figure~\ref{fig:shif-rad}. This indicates that $z_{\rm{g}}$ is dominated by the gravity of the central black hole. In addition, $r_{\rm{g}}/r_{\rm{BLR}}\lesssim 0.01 \ll 1$ and $\langle f\rangle r_{\rm{g}}/r_{\rm{BLR}}< 0.1$ for AGNs in Tables~\ref{sam1}--\ref{sam3}. $\langle f\rangle r_{\rm{g}}/r_{\rm{BLR}}\lesssim 0.01$ for $>$ 96\% AGNs in Sample 1, $>$ 97\% AGNs in Sample 2, and for $>$ 96\% AGNs in Sample 3. Thus, the Schwarzschild metric is valid and matches the weak-field limit at the optical BLR scales of AGNs in our samples, which are conditions on the validity of Equation (1) \citep[see][]{Li22}.

\begin{figure}
  \begin{center}
    \includegraphics[angle=0,scale=0.26]{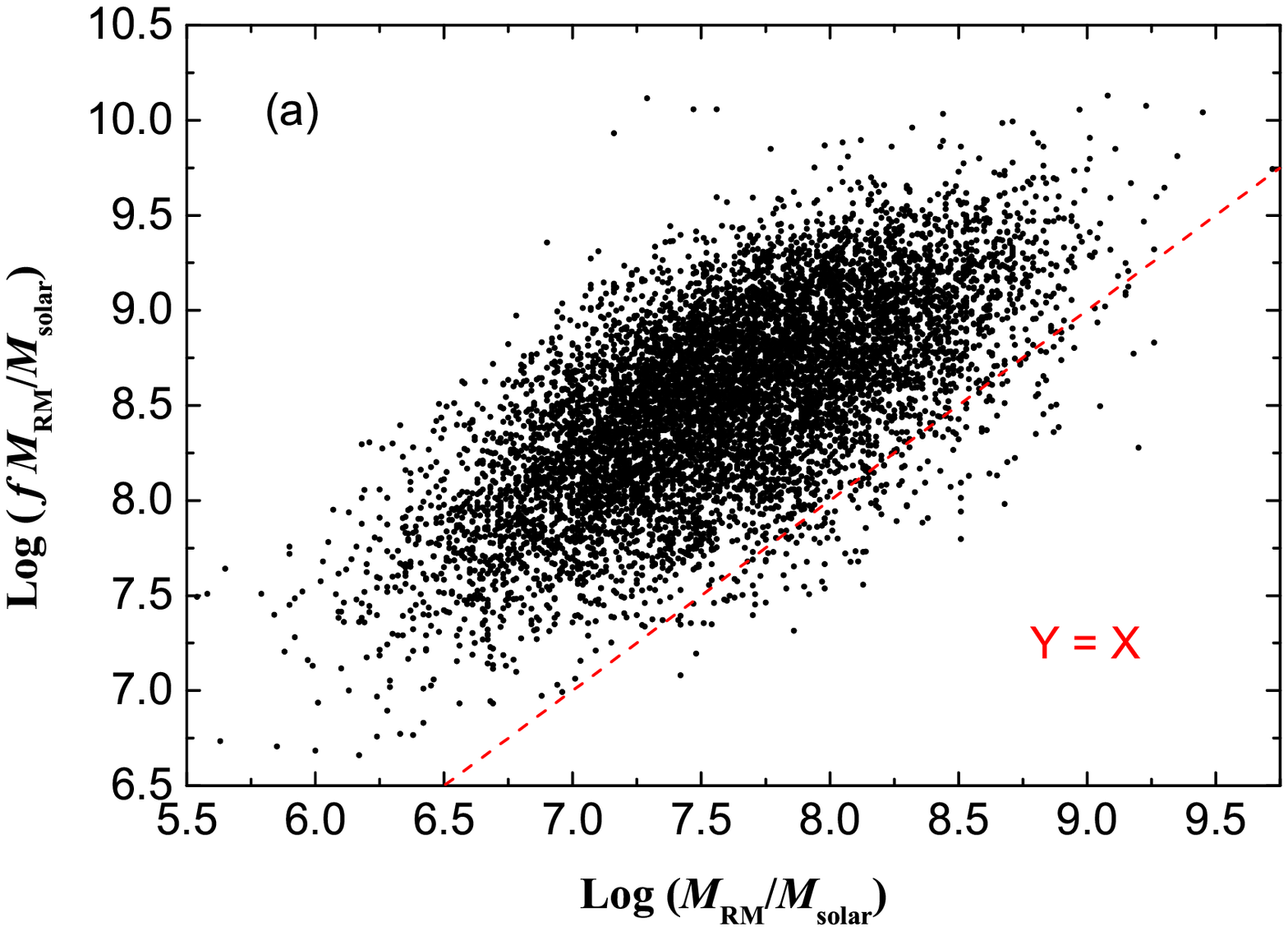}
    \includegraphics[angle=0,scale=0.26]{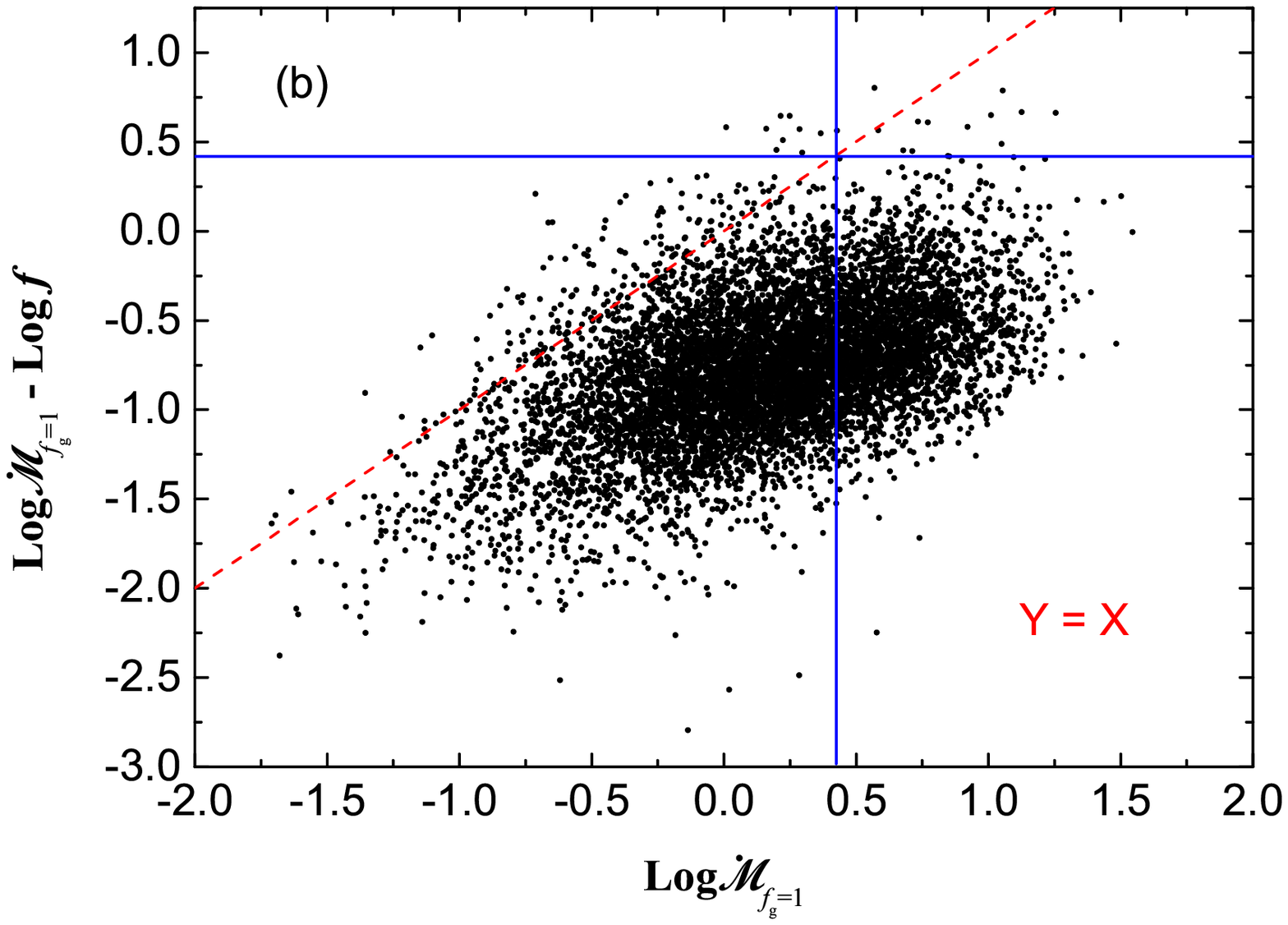}
  \end{center}
  \caption{The $f$ correction effects on mass and accretion rate for 8169 AGNs in Sample 3. The blue lines are $y=0.42$ and $x=0.42$.}
  \label{fig:f-corr}
\end{figure}

Since $f>1$ for most of SDSS AGNs in Samples 1--3, the $f$ correction might result in substantial influence on $M_{\rm{RM}}$ and $\mathscr{\dot M}_{f_{\rm{g}}=1}$. We choose 8169 AGNs in Sample 3 to illustrate this influence. On average, the corrected $M_{\rm{RM}}$ becomes larger by one order of magnitude than $M_{\rm{RM}}$, and the corrected $\mathscr{\dot M}_{f_{\rm{g}}=1}$ decreases by about 10 times (see Figure~\ref{fig:f-corr}). The substantial increase of $M_{\rm{RM}}$ will significantly impact the black hole mass function of these SDSS AGNs, e.g., leading to more AGNs with higher masses. The substantial decrease of $\mathscr{\dot M}_{f_{\rm{g}}=1}$ loosens the requirement for accretion rate of accretion disk, and might make the distinction between high- and low-accreting sources less obvious. If $L_{\rm{bol}}/L_{\rm{Edd}}\geq 0.1$, i.e., $\log \mathscr{\dot M}_{f_{\rm{g}}=1}\geq 0.42$, for high-accreting sources, the percent of AGNs with $\log \mathscr{\dot M}_{f_{\rm{g}}=1}\geq 0.42$ is 31.4\%, but the percent of AGNs with $\log (\mathscr{\dot M}_{f_{\rm{g}}=1}/f) \geq 0.42$ is only 0.3\% (see Figure~\ref{fig:f-corr}). The percent of high-accreting sources is decreased by about 100 times due to the $f$ correction. In a sense, the $f$ correction blurs the distinction between high- and low-accreting sources. 

\begin{figure}
  \begin{center}
    \includegraphics[angle=0,scale=0.15]{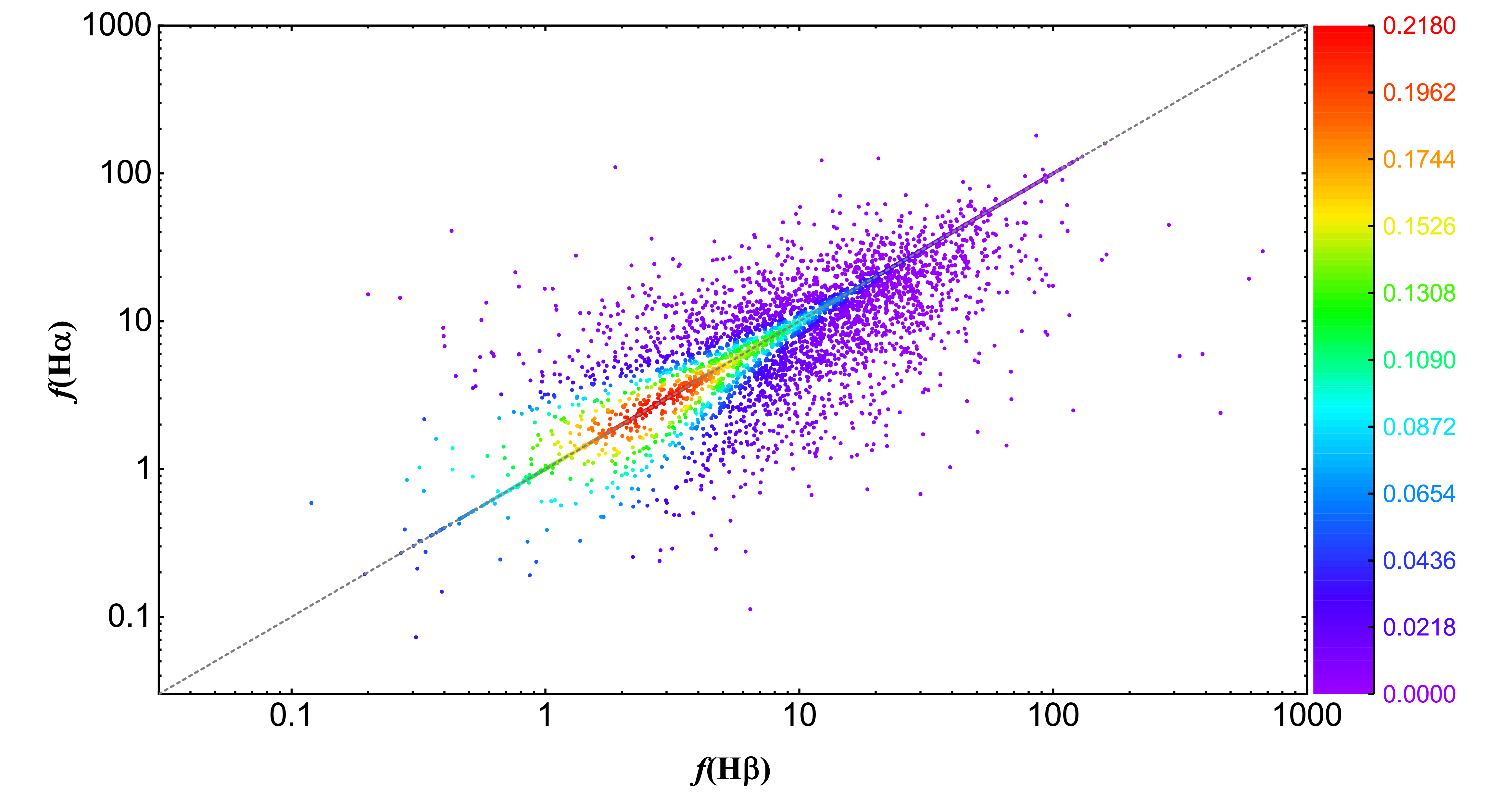}
  \end{center}
  \caption{Density map of $f$(\Ha) vs. $f$(\Hb) for 8169 AGNs in Sample 3. The dashed line is $y=x$.}
  \label{fig:den-hab}
\end{figure}

The virial factor of \Hb\ is consistent with that of \Ha\ for 8169 AGNs in Sample 3 (see Figure~\ref{fig:den-hab}). The \Ha\ lags are consistent with or (slightly) larger than the \Hb\ lags for RM AGNs \citep[e.g.,][]{Ka00,Be10,Gr17}. Because the \Ha\ optical depth is larger than the \Hb\ optical depth, the optical depth effects may result in the larger \Ha\ lags that cause the \Ha\ emission line to seemingly appear at the larger distances than \Hb\ \citep[see][]{Be10}, even though the \Hb\ and \Ha\ broad emission lines are from the same region. Thus, it seems that $r_{\rm{BLR}}$(\Hb) $\approx r_{\rm{BLR}}$(\Ha). For broad emission lines with different $r_{\rm{BLR}}$, there will be $f\propto r_{\rm{BLR}}^{\alpha}$ ($\alpha >0$) as $F_{\rm{r}}$ is considered and the BLR clouds are in the virialized motion for a given AGN \citep{Li17}. The \Hb\ and \Ha\ BLRs have similar virialized kinematics for type-1 AGNs in SDSS DR 16 \citep{Ra22}. If $r_{\rm{BLR}}$(\Hb) $\approx r_{\rm{BLR}}$(\Ha) and $f\propto r_{\rm{BLR}}^{\alpha}$, it will be expected that $f$(\Hb) is on the whole consistent with $f$(\Ha) for AGNs in Sample 3, as shown in Figure~\ref{fig:den-hab}.

\begin{deluxetable}{cccccc}
  \tablecolumns{6}
  \setlength{\tabcolsep}{5pt}
  \tablewidth{0pc}
  \tablecaption{Spearman's rank analysis results}
  \tabletypesize{\scriptsize}
  \tablehead{\colhead{X}  &  \colhead{Y} &  \colhead{Line} & \colhead{$r_{\rm{s}}$} &  \colhead{$P_{\rm{s}}$} &  \colhead{Sample} }

  \startdata

 $\log \mathscr{\dot M}_{f_{\rm{g}}=1}$  & $\log f$ & \Hb\ & 0.615 &  $< 10^{-40}$ & 1 \\

  $\log \mathscr{\dot M}_{f_{\rm{g}}=1}$  & $\log f$ & \Ha\ & 0.578  & $< 10^{-40}$  & 2 \\

  $\log \mathscr{\dot M}_{f_{\rm{g}}=1}$  & $\log f$ & \Hb\ & 0.626  & $< 10^{-40}$  & 3 \\

  $\log \mathscr{\dot M}_{f_{\rm{g}}=1}$ & $\log f$ & \Ha\ & 0.590 & $< 10^{-40}$ & 3 \\

\hline

 $\log L_{\rm{5100}}$ & $\log f$ & \Hb\ & 0.089  & $1.0\times 10^{-17}$  & 1 \\

$\log L_{\rm{5100}}$ & $\log f$ & \Ha\ & 0.047  & $6.0\times 10^{-6}$ & 2 \\

$\log L_{\rm{5100}}$ & $\log f$ & \Hb\ & 0.113  & $8.3\times 10^{-25}$ & 3 \\

$\log L_{\rm{5100}}$ & $\log f$ & \Ha\ & 0.044  & $5.8\times 10^{-5}$ & 3 \\

\hline

$\log [r_{\rm{BLR}}/r_{\rm{g}}]$ & $\log z_{\rm{g}}$ & \Hb\ & -0.440 &  $< 10^{-40}$ & 1  \\

$\log [r_{\rm{BLR}}/r_{\rm{g}}]$ & $\log z_{\rm{g}}$ & \Ha\ & -0.318 &  $< 10^{-40}$ & 2  \\

$\log [r_{\rm{BLR}}/r_{\rm{g}}]$ & $\log z_{\rm{g}}$ & \Hb\ & -0.447 &  $< 10^{-40}$ & 3  \\

$\log [r_{\rm{BLR}}/r_{\rm{g}}]$ & $\log z_{\rm{g}}$ & \Ha\ & -0.335 &  $< 10^{-40}$ & 3  \\

\hline

$\log \mathscr{\dot M}_{f_{\rm{g}}=1}(\eta)$ & $\log f$ & \Hb\ & 0.654 &  $< 10^{-40}$ & 1 \\

$\log \mathscr{\dot M}_{f_{\rm{g}}=1}(\eta)$ & $\log f$ & \Ha\ & 0.647  & $< 10^{-40}$ & 2 \\

$\log \mathscr{\dot M}_{f_{\rm{g}}=1}(\eta)$ & $\log f$ & \Hb\ & 0.662  & $< 10^{-40}$ & 3 \\

$\log \mathscr{\dot M}_{f_{\rm{g}}=1}(\eta)$ & $\log f$ & \Ha\ & 0.658 & $< 10^{-40}$ & 3 \\

\enddata
\tablecomments{\footnotesize X and Y are the relevant quantities presented in Samples 1--3. The first part: $\mathscr{\dot M}_{f_{\rm{g}}=1}=L_{\rm{bol}}/L_{\rm{Edd}}/\eta$ and $\eta=0.038$. The last part: $\mathscr{\dot M}_{f_{\rm{g}}=1}(\eta)=L_{\rm{bol}}/L_{\rm{Edd}}/\eta$ and $\eta=0.089 M_{8}^{0.52}$.}
\label{spearman}
\end{deluxetable}

\begin{deluxetable}{cccccc}
  \tablecolumns{6}
  \setlength{\tabcolsep}{5pt}
  \tablewidth{0pc}
  \tablecaption{Pearson's analysis results}
  \tabletypesize{\scriptsize}
  \tablehead{\colhead{X}  &  \colhead{Y} &  \colhead{Line} & \colhead{$r$} &  \colhead{$P$} &  \colhead{Sample} }

  \startdata

 $\log \mathscr{\dot M}_{f_{\rm{g}}=1}$ & $\log f$ & \Hb\ & 0.601 &  $< 10^{-40}$ & 1 \\

 $\log \mathscr{\dot M}_{f_{\rm{g}}=1}$  & $\log f$& \Ha\ & 0.568  & $< 10^{-40}$  & 2 \\

 $\log \mathscr{\dot M}_{f_{\rm{g}}=1}$  & $\log f$ & \Hb\ & 0.618  & $< 10^{-40}$  & 3 \\

$\log \mathscr{\dot M}_{f_{\rm{g}}=1}$ & $\log f$ & \Ha\ & 0.588  & $< 10^{-40}$ & 3 \\

\hline

 $\log \mathscr{\dot M}_{f_{\rm{g}}=1}$ & $\log L_{\rm{5100}}$ & \Hb\ & 0.236 &  $< 10^{-40}$
 & 1 \\

 $\log \mathscr{\dot M}_{f_{\rm{g}}=1}$  & $\log L_{\rm{5100}}$ & \Ha\ & 0.265  & $< 10^{-40}$      & 2 \\

 $\log \mathscr{\dot M}_{f_{\rm{g}}=1}$ & $\log L_{\rm{5100}}$ & \Hb\ & 0.247  & $< 10^{-40}$      & 3 \\

$\log \mathscr{\dot M}_{f_{\rm{g}}=1}$ & $\log L_{\rm{5100}}$ & \Ha\ & 0.247 & $< 10^{-40}$
 & 3 \\

\hline

 $\log \mathscr{\dot M}_{f_{\rm{g}}=1}$  & $\log v_{\rm{FWHM}}$ & \Hb\ & -0.747 &  $< 10^{-40}$
 & 1 \\

 $\log \mathscr{\dot M}_{f_{\rm{g}}=1}$ & $\log v_{\rm{FWHM}}$ & \Ha\ & -0.757 & $< 10^{-40}$      & 2 \\

 $\log \mathscr{\dot M}_{f_{\rm{g}}=1}$  & $\log v_{\rm{FWHM}}$ & \Hb\ & -0.753 & $< 10^{-40}$      & 3 \\

$\log \mathscr{\dot M}_{f_{\rm{g}}=1}$  & $\log v_{\rm{FWHM}}$ & \Ha\ & -0.787 & $< 10^{-40}$      & 3 \\

\hline

 $\log L_{\rm{5100}}$ & $\log f$ & \Hb\ & 0.072  & $5.4\times 10^{-12}$  & 1 \\

$\log L_{\rm{5100}}$ & $\log f$ & \Ha\ & 0.032  & $2.1\times 10^{-3}$ & 2 \\

$\log L_{\rm{5100}}$ & $\log f$ & \Hb\ & 0.096  & $4.6\times 10^{-18}$ & 3 \\

$\log L_{\rm{5100}}$ & $\log f$ & \Ha\ & 0.028  & $1.0\times 10^{-2}$ & 3 \\

\hline

$\log L_{\rm{5100}}$ & $\log v_{\rm{FWHM}}$ & \Hb\ & 0.133  & $1.5\times 10^{-37}$  & 1 \\

$\log L_{\rm{5100}}$ & $\log v_{\rm{FWHM}}$ & \Ha\ & 0.073  & $2.7\times 10^{-12}$ & 2 \\

$\log L_{\rm{5100}}$ & $\log v_{\rm{FWHM}}$ & \Hb\ & 0.117  & $3.9\times 10^{-26}$ & 3 \\

$\log L_{\rm{5100}}$ & $\log v_{\rm{FWHM}}$ & \Ha\ & 0.090  & $3.3\times 10^{-16}$ & 3 \\

\hline

$\log v_{\rm{FWHM}}$ & $\log f$ & \Hb\ & -0.647 &  $< 10^{-40}$ & 1  \\

$\log v_{\rm{FWHM}}$ & $\log f$ & \Ha\ & -0.686 &  $< 10^{-40}$ & 2  \\

$\log v_{\rm{FWHM}}$ & $\log f$ & \Hb\ & -0.666 &  $< 10^{-40}$ & 3  \\

$\log v_{\rm{FWHM}}$ & $\log f$ & \Ha\ & -0.693 &  $< 10^{-40}$ & 3  \\

\enddata
\tablecomments{\footnotesize X and Y are the relevant quantities presented in Samples 1--3.  $\mathscr{\dot M}_{f_{\rm{g}}=1}=L_{\rm{bol}}/L_{\rm{Edd}}/\eta$ and $\eta=0.038$.}
\label{pearson}
\end{deluxetable}

\begin{deluxetable}{cccccc}
  \tablecolumns{6}
  \setlength{\tabcolsep}{5pt}
  \tablewidth{0pc}
  \tablecaption{Partial correlation analysis results}
  \tabletypesize{\scriptsize}
  \tablehead{\colhead{Name}  & \colhead{Order}  & \colhead{Line} & \colhead{$r_{\rm{p}}$} &  \colhead{$P_{\rm{p}}$} &  \colhead{Sample} }

  \startdata

 $r_{\log f \log \mathscr{\dot M}_{f_{\rm{g}}=1},\log L_{\rm{5100}}}$ & 1 & \Hb\ & 0.603 &
 $< 10^{-4}$ & 1 \\

 $r_{\log f \log \mathscr{\dot M}_{f_{\rm{g}}=1},\log v_{\rm{FWHM}}}$ & 1 & \Hb\ &  0.232  &
   $< 10^{-4}$    & 1 \\

 $r_{\log f \log \mathscr{\dot M}_{f_{\rm{g}}=1},\log v_{\rm{FWHM}} \log L_{\rm{5100}}}$ & 2 & \Hb\ & 0.151 & $< 10^{-4}$    & 1 \\

\hline

 $r_{\log f \log \mathscr{\dot M}_{f_{\rm{g}}=1},\log L_{\rm{5100}}}$ & 1 & \Ha\ &  0.581 &
 $< 10^{-4}$ & 2 \\

   $r_{\log f \log \mathscr{\dot M}_{f_{\rm{g}}=1},\log v_{\rm{FWHM}}}$ & 1 & \Ha\ &  0.102 &
   $< 10^{-4}$    & 2 \\

   $r_{\log f \log \mathscr{\dot M}_{f_{\rm{g}}=1},\log v_{\rm{FWHM}} \log L_{\rm{5100}}}$ & 2 & \Ha\ & 0.055 & $< 10^{-4}$   & 2 \\

\hline

 $r_{\log f \log \mathscr{\dot M}_{f_{\rm{g}}=1},\log L_{\rm{5100}}}$ & 1 & \Hb\ & 0.616 &
 $< 10^{-4}$ & 3 \\

   $r_{\log f \log \mathscr{\dot M}_{f_{\rm{g}}=1},\log v_{\rm{FWHM}}}$ & 1 & \Hb\ &  0.237 &
   $< 10^{-4}$    & 3 \\

   $r_{\log f \log \mathscr{\dot M}_{f_{\rm{g}}=1},\log v_{\rm{FWHM}} \log L_{\rm{5100}}}$ & 2 & \Hb\ &  0.141 & $< 10^{-4}$   & 3 \\

\hline

$r_{\log f \log \mathscr{\dot M}_{f_{\rm{g}}=1},\log L_{\rm{5100}}}$ & 1 & \Ha\ & 0.600 &
$< 10^{-4}$ & 3 \\

$r_{\log f \log \mathscr{\dot M}_{f_{\rm{g}}=1},\log v_{\rm{FWHM}}}$ & 1 & \Ha\ & 0.096 &
$< 10^{-4}$    & 3 \\

$r_{\log f \log \mathscr{\dot M}_{f_{\rm{g}}=1},\log v_{\rm{FWHM}} \log L_{\rm{5100}}}$ & 2
& \Ha\ & 0.035 & $1.6\times 10^{-3}$   & 3 \\

\enddata
\tablecomments{\footnotesize Based on the Pearson's correlation coefficient $r$ in Table~\ref{pearson}, the partial correlation coefficient $r_{\rm{p}}$ and the p-value $P_{\rm{p}}$ of the hypothesis test are estimated using the Website for Statistical Computation (http://vassarstats.net/index.html).  Orders 1 and 2 denote the 1st and 2nd order partial correlation coefficients, respectively. $\mathscr{\dot M}_{f_{\rm{g}}=1}=L_{\rm{bol}}/L_{\rm{Edd}}/\eta$ and $\eta=0.038$.}
\label{partial}
\end{deluxetable}

\section{POTENTIAL INFLUENCE ON QUASARS AT $z\gtrsim 6$}\label{sec:quas}
Quasars at $z\gtrsim 6$ can probe the formation and growth of SMBHs in the Universe within the first billion years after the Big Bang. The quasars, with $M_{\rm{RM}}\gtrsim 10^{9} M_{\odot}$ at $z\gtrsim 7$ and with $M_{\rm{RM}}\gtrsim 10^{10} M_{\odot}$ at $z\gtrsim 6$, make the formation and growth of SMBHs ever more challenging \citep[e.g.,][]{Wu15,Fa23}. These SMBHs will need a combination of massive early black hole seeds with highly efficient and sustained accretion \citep[e.g.,][]{Fa23}. However, the single-epoch spectrum method had been widely used to estimate $M_{\rm{RM}}$ of the high-$z$ quasars \citep[e.g.,][]{Wi10,Wu15,Wa19,Ei23}, and this may result in underestimation of $M_{\rm{RM}}$, overestimation of $L_{\rm{bol}}/L_{\rm{Edd}}$, and significant influence on the formation and growth of SMBHs in the early Universe. Based on Equation (4), We will use $\log f = 0.8+0.8\log \mathscr{\dot M}_{f_{\rm{g}}=1}$ to estimate $f$ and study its influence for quasars at $z\gtrsim 6$.

There are 113 quasars at $6 \lesssim z \lesssim 8$ with reliable Mg~{\sc ii}-based black hole mass estimates \citep{Fa23}. These 113 quasars have $\langle f\rangle =78$ ($f=$ 12--189), $\langle \log (M_{\rm{RM}}/M_{\odot})\rangle= 9.0$, $\langle \log (f M_{\rm{RM}}/M_{\odot})\rangle= 10.9$, and $\langle L_{\rm{bol}}/L_{\rm{Edd}}/f\rangle= 0.01$ (see Table~\ref{highquasar}). The $f$ correction makes $M_{\rm{RM}}$ increase by one--two orders of magnitude. Also, substantially reduced $L_{\rm{bol}}/L_{\rm{Edd}}/f=$ 0.007--0.014 will make these 113 quasars accreting at well below the Eddington limit, although likely in the radiatively efficient regime via a geometrically thin, optically thick accretion disk \citep{Sh73}. Based on Equation (7) in \citet{Fa23}, $M_{\bullet}(t)\propto \exp(t)$, the growth times of SMBHs in these quasars will increase by a factor of 2.5--5.2 due to the $f$ correction. Thus, the black hole seeds don't seem to have enough time to grow up for SMBHs in quasars at $z\gtrsim 6$, and this gives more strong constraints on the formation and growth of the black hole seeds. Thus, the $f$ correction will make it more difficult to explain the formation and growth of SMBHs at $z\gtrsim 6$, e.g., larger masses of SMBHs need more massive early black hole seeds and/or longer growth times. \citet{Bo23} found evidence for heavy-seed origin of early SMBHs from a $z\approx$10 X-ray quasar. Our results of corrected masses support heavy-seed origin scenarios of early SMBHs.

The highest redshift quasar J031343.84-180636.40, at $z=7.6423$, has $\log(M_{\rm{RM}}/M_{\odot})=9.2$, $\log(fM_{\rm{RM}}/M_{\odot})=11.0$, and $L_{\rm{bol}}/L_{\rm{Edd}}/f=0.01$. Quasar J0100+2802, the most luminous quasar known at $z > 6$, has $\log (M_{\rm{RM}}/M_{\odot})\sim 10$, and $L_{\rm{bol}}/L_{\rm{Edd}} \sim 0.8$ \citep{Wu15}. So, J0100+2802 has $\log (fM_{\rm{RM}}/M_{\odot})\sim 12$, and $L_{\rm{bol}}/L_{\rm{Edd}}/f \sim 0.01$. Quasar J140821.67+025733.2, at $z=2.055$, has a black hole mass of $10^{11.3}M_{\odot}$ with the uncertainty of 0.4 dex \citep{Kz17}. It seems reasonable that J0100+2802 has $\log (fM_{\rm{RM}}/M_{\odot})\sim 12$ with the uncertainty of 0.4. Recently, \citet{Ko23} found an AGN at $z=8.50$ with an \Hb-based mass of $\log (M_{\rm{RM}}/M_{\odot})=8.2$, and $L_{\rm{bol}}/L_{\rm{Edd}}\sim 0.33$. We have $\log (fM_{\rm{RM}}/M_{\odot})\sim 9.7$, and $L_{\rm{bol}}/L_{\rm{Edd}}/f \sim 0.01$ for this AGN. In addition, Samples 1--3 each have $\langle L_{\rm{bol}}/L_{\rm{Edd}}/f\rangle= 0.01$. It is interesting that $\langle L_{\rm{bol}}/L_{\rm{Edd}}/f\rangle= 0.01$ exists for quasars/AGNs at $z < 0.35$ and $z\gtrsim 6$, and perhaps it is a coincidence.

\begin{deluxetable}{cccccccc}
  \tablecolumns{8}
  \setlength{\tabcolsep}{5pt}
  \tablewidth{0pc}
  \tablecaption{Quasars at $z\gtrsim 6$}
  \tabletypesize{\scriptsize}
  \tablehead{\colhead{Name}  & \colhead{redshift}  & \colhead{$\log \frac{M_{\rm{RM}}}{M_{\odot}}$}& \colhead{$\log \frac{L_{\rm{3000}}}{\rm{erg~s^{-1}}}$} & \colhead{$ \frac{L_{\rm{bol}}}{L_{\rm{Edd}}}$} &  \colhead{$f$} &  \colhead{$\log \frac{fM_{\rm{RM}}}{M_{\odot}}$} & \colhead{$ \frac{L_{\rm{bol}}}{f L_{\rm{Edd}}}$}\\
  \colhead{(1)}  &\colhead{(2)}  &\colhead{(3)}  &\colhead{(4)}  &\colhead{(5)}  &\colhead{(6)}  &  \colhead{(7)}&  \colhead{(8)}}

\startdata

J000825.77-062604.42&5.930&	8.72& 46.32&1.626&	127.4& 10.82 &0.013  \\
J002031.47-365341.82&6.834&	9.23& 46.42&0.633&	59.8& 11.01& 0.011\\
J002429.77+391318.97&6.621& 8.43& 46.18& 2.293&167.7&10.66& 0.014\\
...  &...  &...  &...	&...	&...	&... &...	\\

\enddata
\tablecomments{\footnotesize $L_{\rm{bol}}=5.15 L_{3000}$, and $L_{3000}$ is the UV quasar continuum luminosity at rest-frame wavelength 3000 \AA\ \citep{Fa23}. $f$ is estimated by $\log f = 0.8+0.8\log \mathscr{\dot M}_{f_{\rm{g}}=1}$, where $\mathscr{\dot M}_{f_{\rm{g}}=1}$ is estimated by $L_{\rm{bol}}/L_{\rm{Edd}}/\eta$ and $\eta=0.038$.\\(This table is available in its entirety in machine-readable form.)}
\label{highquasar}
\end{deluxetable}

\section{POTENTIAL INFLUENCE ON $M_{\bullet}-\sigma_{\ast}$ MAP OF AGNs}\label{sec:msig}
It is not determined whether or not the SMBHs coevolve with their host galaxies \citep[e.g.,][]{Ko13}, especially, the SMBHs in AGNs with high accrete rates. Coevolution had been supported by the $M_{\bullet}-\sigma_{\ast}$ relations of local quiescent galaxies. For 31 nearby galaxies, \citet{Tr02} obtained $\log(M_{\bullet}/M_{\odot}) = 8.13+4.02 \log(\sigma_{\ast}/\sigma_0)$ with $\sigma_0=200 ~\rm{km~s^{-1}}$. \citet{Mc13} presented a revised scaling relation of $\log(M_{\bullet}/M_{\odot}) = 8.32 + 5.64 \log(\sigma_{\ast}/\sigma_0)$ for dynamical measurements of $M_{\bullet}$ at the centers of 72 nearby galaxies. For 72 nearby quiescent galaxies with dynamical measurements of $M_{\bullet}$, \citet{Wo13} obtained $\log(M_{\bullet}/M_{\odot}) = 8.37 +5.31 \log(\sigma_{\ast}/\sigma_0)$.  For 19 local luminous AGNs at $z < 0.01$, \citet{Ca20} obtained $\log(M_{\bullet}/M_{\odot}) = 8.14 + 3.38 \log(\sigma_{\ast}/\sigma_0)$.

\begin{figure}
  \begin{center}
    \includegraphics[angle=0,scale=0.35]{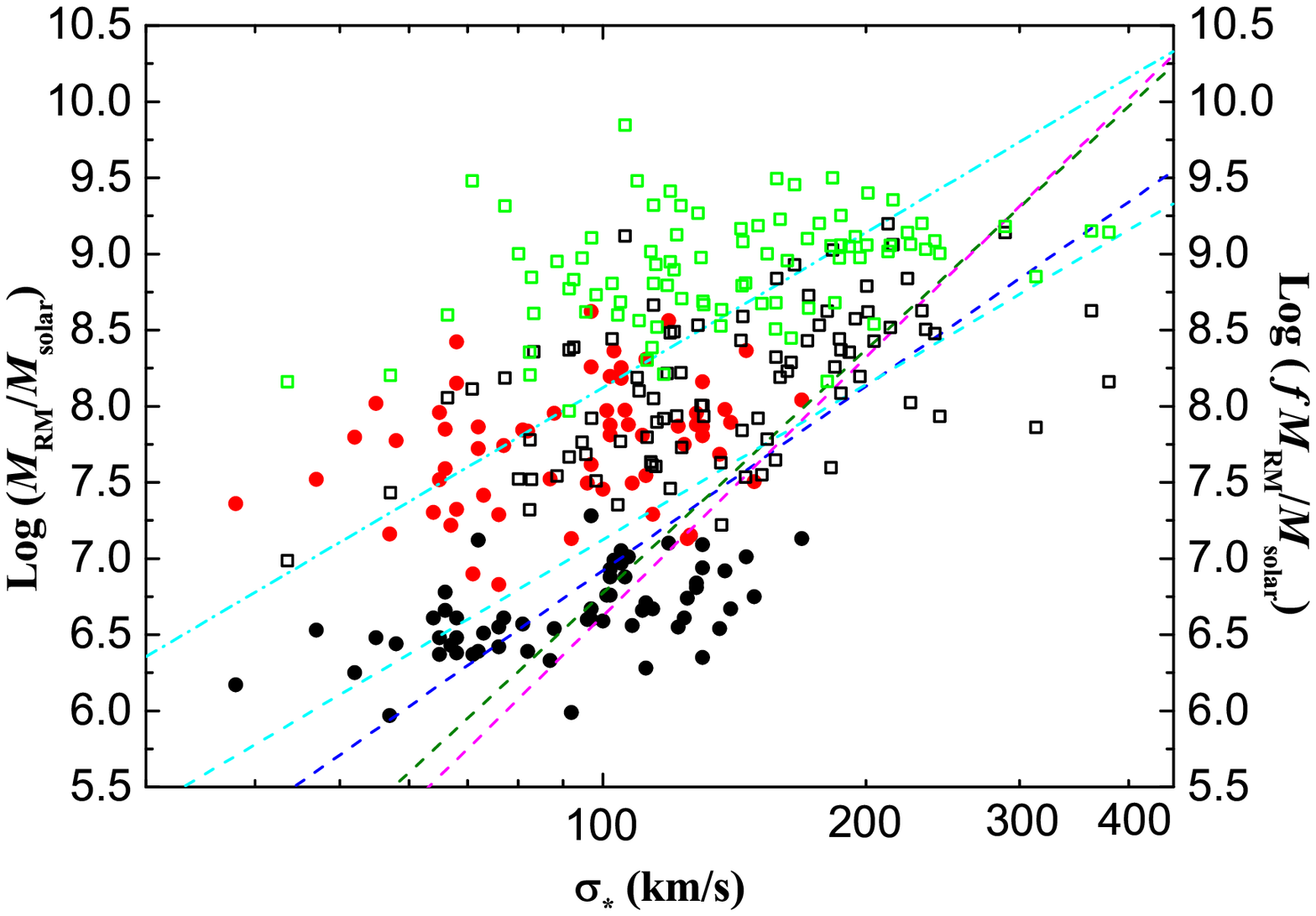}
  \end{center}
  \caption{$M_{\bullet}-\sigma_{\ast}$ map for 62 AGNs in our samples (solid circles), and 88 quasars in \citet{Sh15b} (open squares). The black symbols correspond to $M_{\rm{RM}}$, and the colourful symbols are $fM_{\rm{RM}}$. The bule dashed line is the \citet{Tr02} relation for nearby inactive galaxies. The olive dashed line is the \citet{Wo13} relation for nearby quiescent galaxies. The magenta dashed line is the \citet{Mc13} relation for 72 nearby galaxies. The cyan dashed line is the \citet{Ca20} relation for local luminous AGNs. The cyan dash-dotted line is the \citet{Ca20} relation moved vertically upward by an order of magnitude.}
  \label{fig:m-sigma}
\end{figure}

We collect $\sigma_{\ast}$ from Table 1 in \citet{Wo15} for AGNs in our samples, and $M_{\rm{RM}}$, $\sigma_{\ast}$, $L_{\rm{5100}}$, etc for other SDSS quasars from Table 1
in \citet{Sh15b}. There are 62 AGNs and 88 quasars collected (see Tables~\ref{M-sigma62}--\ref{M-sigma88}). These 62 AGNs are at $z=$ 0.013--0.100 with $\langle z\rangle =$ 0.063, which are beyond the local Universe. These 88 quasars are at
$z=$ 0.116--0.997 with $\langle z\rangle =$ 0.581, which are well beyond the local Universe. The significant difference of redshift might influence whether these 150 sources follow
the same $M_{\bullet}-\sigma_{\ast}$ relationship as the local sources. The ($M_{\rm{RM}}$,$\sigma_{\ast}$) data of these 150 soures do not roughly follow these four local $M_{\bullet}-\sigma_{\ast}$ relations, and the deviation of these 88 quasars is more obvious (see Figure~\ref{fig:m-sigma}). As study the coevolution of SMBHs with host galaxies, the local $M_{\bullet}-\sigma_{\ast}$ relation is basically equivalent to the local black hole--galaxy bulge mass relation. Quasars at $z\sim 6$ are above the local mass relation \citep[e.g.,][]{Fa23}. So, these quasars at $z\sim 6$ should be above the local $M_{\bullet}-\sigma_{\ast}$ relation. Thus, these $z\gtrsim 6$ quasars using $fM_{\rm{RM}}$ might be above these local $M_{\bullet}-\sigma_{\ast}$ relations.

These local luminous AGNs in \citet{Ca20} have $\langle L_{\rm{bol}}/L_{\rm{Edd}}\rangle= 0.07$, which corresponds to $f=10$, indicating that the $M_{\bullet}-\sigma_{\ast}$ relation of \citet{Ca20} should be corrected by moving vertically upward by an order of magnitude in the $M_{\bullet}-\sigma_{\ast}$ map. Though, the ($fM_{\rm{RM}}$,$\sigma_{\ast}$) data of these 150 sources deviate from (are above) these local $M_{\bullet}-\sigma_{\ast}$ relations, they roughly follow the corrected $M_{\bullet}-\sigma_{\ast}$ relation (see Figure~\ref{fig:m-sigma}). This deviation implies requirements of more massive black hole seeds, longer growth times, larger AGN duty cycles, and/or higher mass accretion rates in long-term accretion history for them. Also, it seems that agreement of the ($fM_{\rm{RM}}$,$\sigma_{\ast}$) data with the corrected $M_{\bullet}-\sigma_{\ast}$ relation is better than agreement of the ($M_{\rm{RM}}$,$\sigma_{\ast}$) data with these local $M_{\bullet}-\sigma_{\ast}$ relations (see Figure~\ref{fig:m-sigma}). These results might shed light on possible redshift evolution in the $M_{\bullet}-\sigma_{\ast}$ relationship. The formation and growth of the local SMBHs and host galaxies might be different from those of the SMBHs in higher redshift AGNs/quasars and host galaxies.

\begin{deluxetable}{ccccccc}
  \tablecolumns{7}
  \setlength{\tabcolsep}{5pt}
  \tablewidth{0pc}
  \tablecaption{62 SDSS AGNs in $M_{\bullet}-\sigma_{\ast}$ map research}
  \tabletypesize{\scriptsize}
  \tablehead{\colhead{Name}  & \colhead{redshift}  & \colhead{$\log \frac{M_{\rm{RM}}}{M_{\odot}}$} & \colhead{$ \frac{L_{\rm{bol}}}{L_{\rm{Edd}}}$} &  \colhead{$f$} & \colhead{$\frac{\sigma_{\ast}}{\rm{km~s^{-1}}}$} & \colhead{$\log \frac{fM_{\rm{RM}}}{M_{\odot}}$}\\
  \colhead{(1)}  &\colhead{(2)}  &\colhead{(3)}  &\colhead{(4)}  &\colhead{(5)}  &\colhead{(6)}  &  \colhead{(7)} }

\startdata

J010409.16+000843.7&0.071&6.78&	0.065&6.4&66$\pm$16&7.59   \\
J030417.78+002827.4&0.045&6.54&	0.147&25.9&	88$\pm$8&7.95   \\
J073106.87+392644.7&0.048&6.39&	0.120&29.8&	72$\pm$14&7.86  \\
...  &...  &...  &...	&...	&... &	... \\

\enddata
\tablecomments{\footnotesize $\sigma_{\ast}$ of 62 AGNs in our samples are taken from Table 1 in \citet{Wo15}. \\(This table is available in its entirety in machine-readable form.)}
\label{M-sigma62}
\end{deluxetable}

\begin{deluxetable}{cccccccc}
  \tablecolumns{8}
  \setlength{\tabcolsep}{5pt}
  \tablewidth{0pc}
  \tablecaption{88 SDSS quasars in $M_{\bullet}-\sigma_{\ast}$ map research}
  \tabletypesize{\scriptsize}
  \tablehead{\colhead{Name}  & \colhead{redshift}  & \colhead{$\log \frac{M_{\rm{RM}}}{M_{\odot}}$} & \colhead{$\log \frac{L_{\rm{5100}}}{\rm{erg~s^{-1}}}$} &\colhead{$ \frac{L_{\rm{bol}}}{L_{\rm{Edd}}}$} &  \colhead{$f$} & \colhead{$\frac{\sigma_{\ast}}{\rm{km~s^{-1}}}$} & \colhead{$\log \frac{fM_{\rm{RM}}}{M_{\odot}}$}\\
  \colhead{(1)}  &\colhead{(2)}  &\colhead{(3)}  &\colhead{(4)}  &\colhead{(5)}  &\colhead{(6)}  &  \colhead{(7)}  &\colhead{(8)} }

\startdata

141359.51+531049.3&0.8982&7.94&44.11&0.117&15.6&121$\pm$31&9.13 \\
141324.28+530527.0&0.4559&8.36&43.91&0.028&	4.9&191$\pm$4&9.05  \\
141323.27+531034.3&0.8492&8.93&44.28&0.017&3.4&	166$\pm$20&9.46 \\
...  &...  &...  &...	&... &... &... &	... \\

\enddata
\tablecomments{\footnotesize 88 SDSS quasars are taken from Table 1 in \citet{Sh15b}. $L_{\rm{bol}}=9.8 L_{\rm{5100}}$. $f$ is estimated by $\log f = 0.8+0.8\log \mathscr{\dot M}_{f_{\rm{g}}=1}$, where $\mathscr{\dot M}_{f_{\rm{g}}=1}$ is estimated by $L_{\rm{bol}}/L_{\rm{Edd}}/\eta$ and $\eta=0.038$.\\(This table is available in its entirety in machine-readable form.)}
\label{M-sigma88}
\end{deluxetable}

\section{DISCUSSION}\label{sec:diss}
\begin{figure}
  \begin{center}
   \includegraphics[angle=0,scale=0.10]{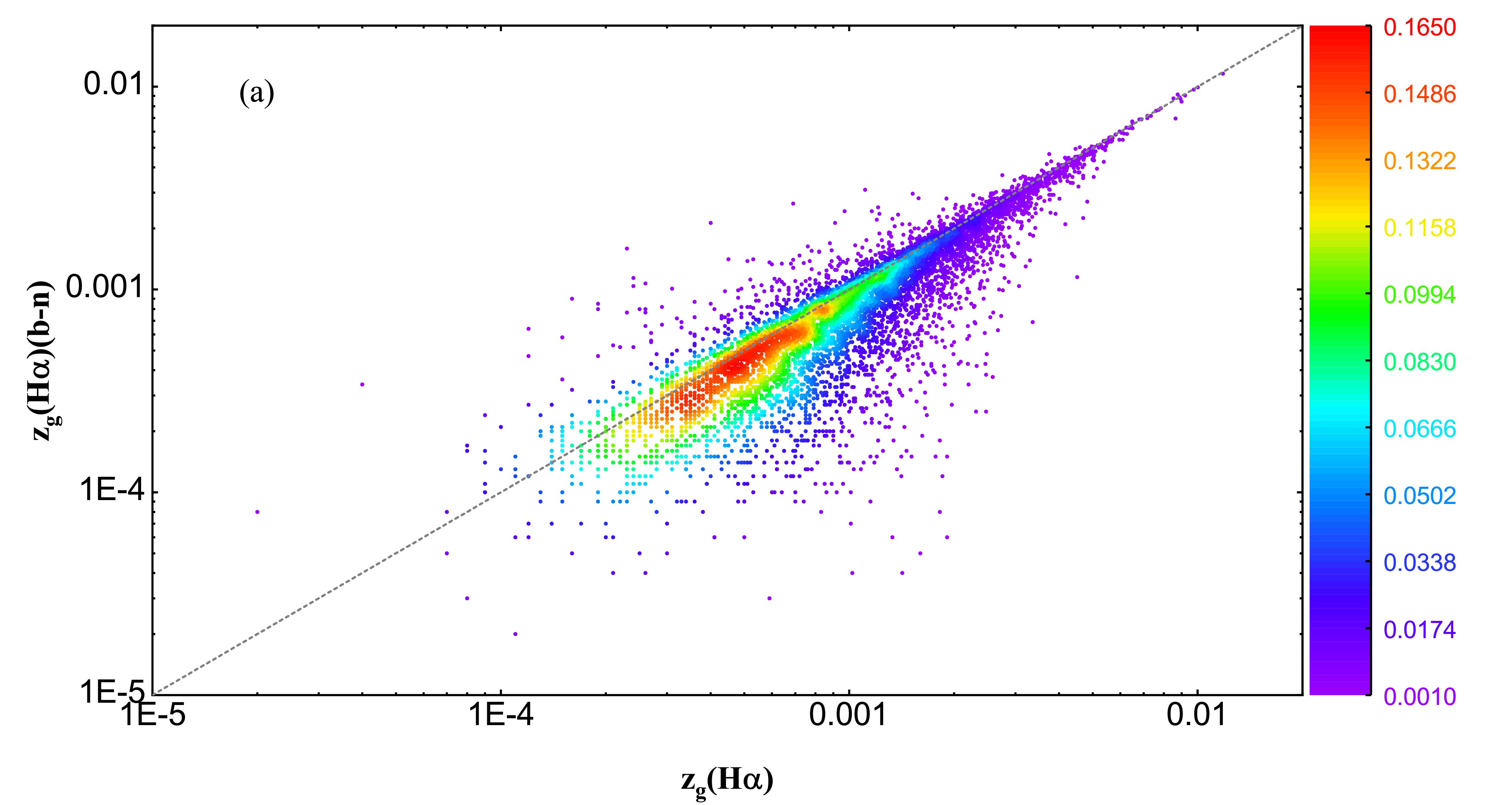}
   \includegraphics[angle=0,scale=0.10]{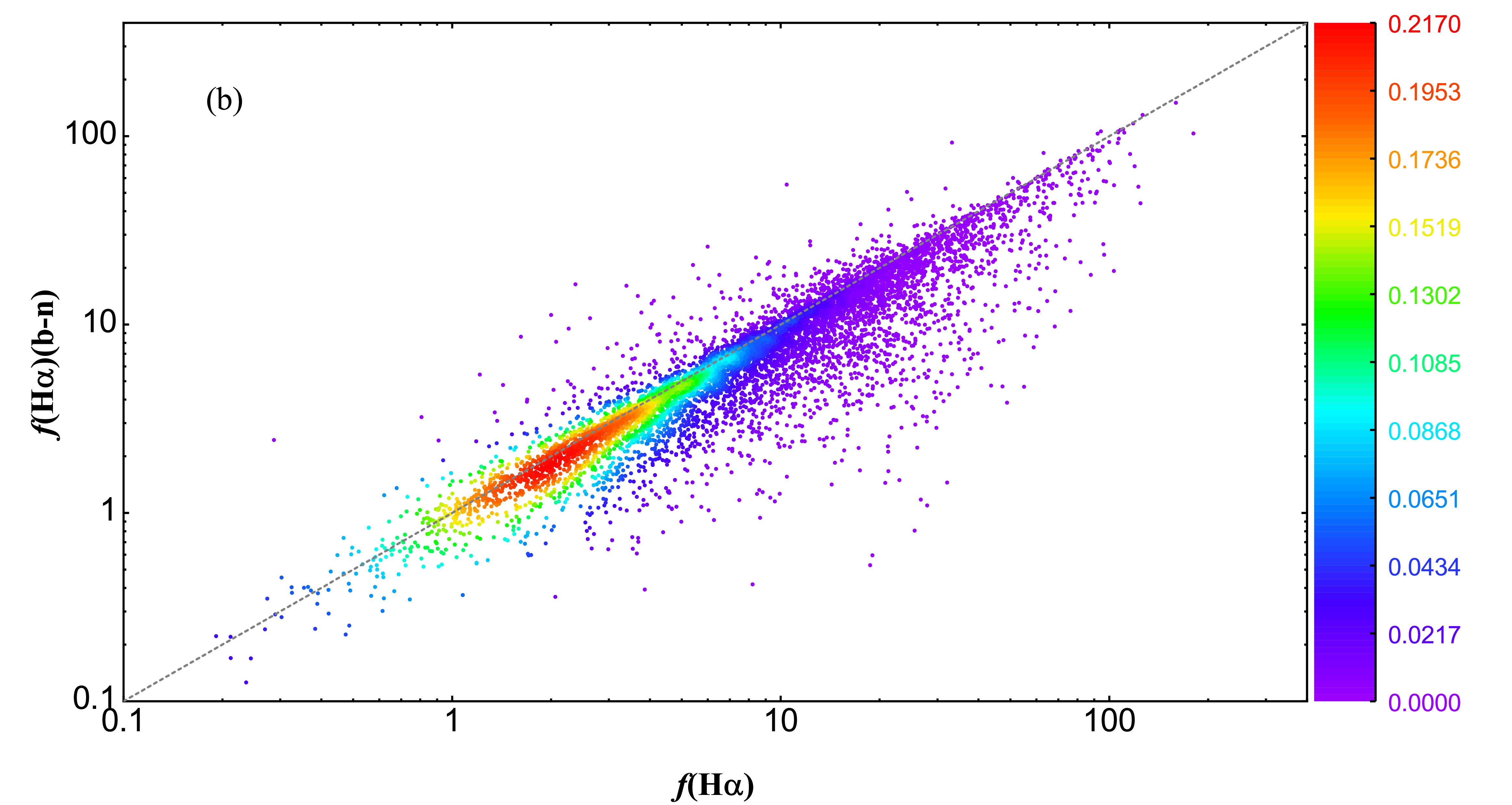}
  \end{center}
  \caption{Density maps for 7552 AGNs in Sample 2. Panel ($a$): narrow \Ha-based $z_{\rm{g}}$(\Ha)(b-n) vs. \oiii$\lambda$5007-based $z_{\rm{g}}$(\Ha). Panel ($b$): $f$(\Ha)(b-n) vs. $f$(\Ha). The dashed lines are $y=x$.}
  \label{fig:den-sam2}
\end{figure}

The redward shift $z_{\rm{g}}$ can also be estimated by $\lambda_{\rm{b}}$ and $\lambda_{\rm{n}}$ of the \Hb\ and \Ha\ lines (see Equation 3). Because of the absence of the uncertainty of $\lambda_{\rm{n}}$ for \Hb\ in Table 2 of \citet{Li19}, $z_{\rm{g}}$ is estimated by $\lambda_{\rm{b}}$ and $\lambda_{\rm{n}}$ of \Ha\ for 7552 AGNs in Sample 2, $z_{\rm{g}}$(\Ha)(b-n). $z_{\rm{g}}$(\Ha)(b-n) is roughly \textbf{consistent} with \oiii$\lambda$5007-based $z_{\rm{g}}$(\Ha) (see Figure~\ref{fig:den-sam2}). Considering the uncertainties of $z_{\rm{g}}$(\Ha) and $z_{\rm{g}}$(\Ha)(b-n) (see columns 3--4 in Table~\ref{sam2}), they are consistent with each other. Thus, the results of $z_{\rm{g}}$(\Ha) are reliable. Based on $z_{\rm{g}}$(\Ha) and $z_{\rm{g}}$(\Ha)(b-n), the virial factors of $f$(\Ha) and $f$(\Ha)(b-n) are estimated and are compared to test their reliabilities. Figure~\ref{fig:den-sam2} shows that $f$(\Ha) and $f$(\Ha)(b-n) are basically consistent. Considering the uncertainties, which have a mean of 2.0 and a median of 0.8 for $f$(\Ha) and a mean of 1.6 and a median of 0.7 for $f$(\Ha)(b-n) (see columns 8--9 in Table~\ref{sam2}), $f$(\Ha) is consistent with $f$(\Ha)(b-n). Thus, the selection of \oiii$\lambda$5007 as a reference to estimate $z_{\rm{g}}$ in Equation (3) will not influence our results.

\begin{figure}
  \begin{center}
   \includegraphics[angle=0,scale=0.26]{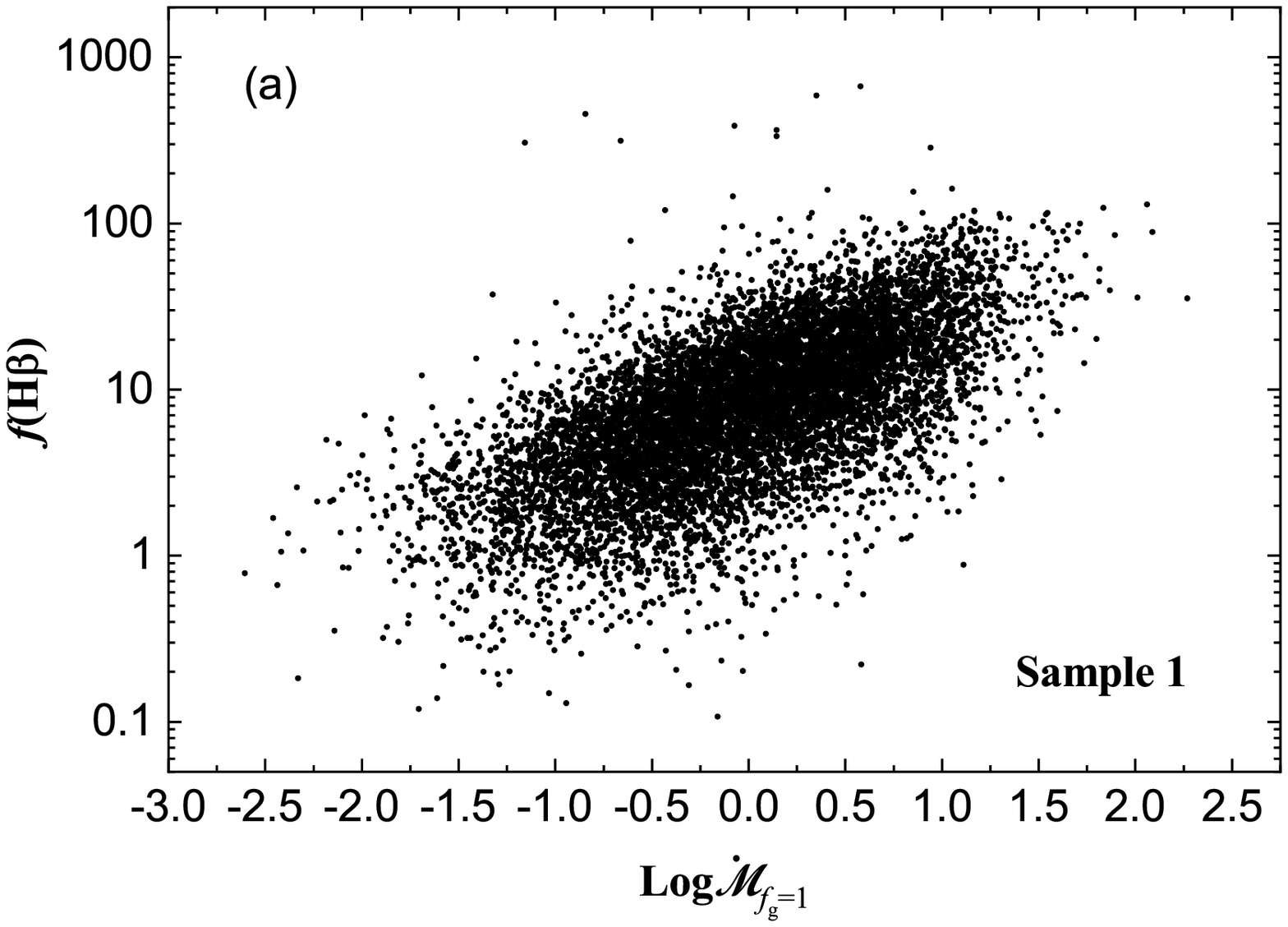}
   \includegraphics[angle=0,scale=0.26]{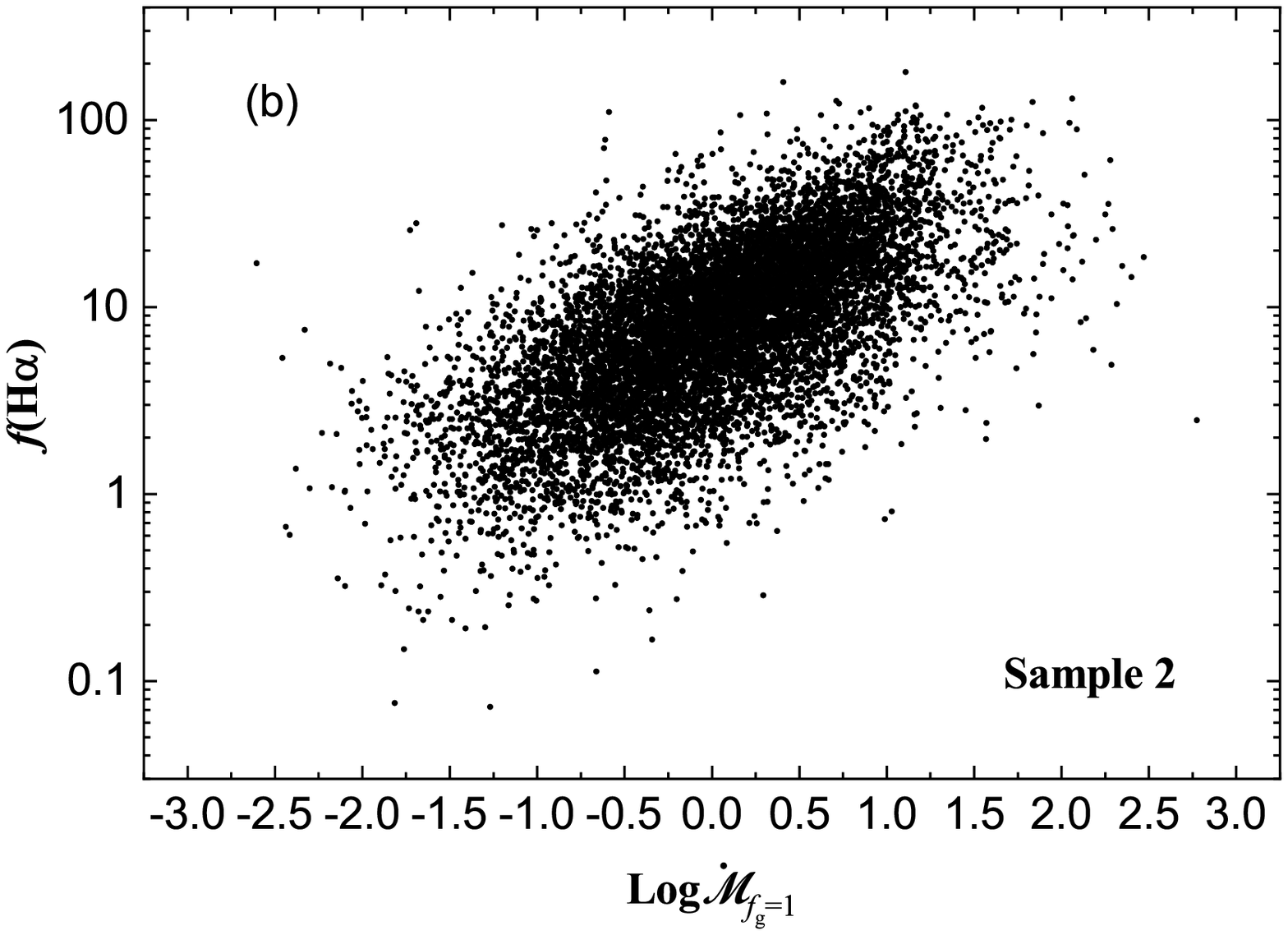}
   \includegraphics[angle=0,scale=0.26]{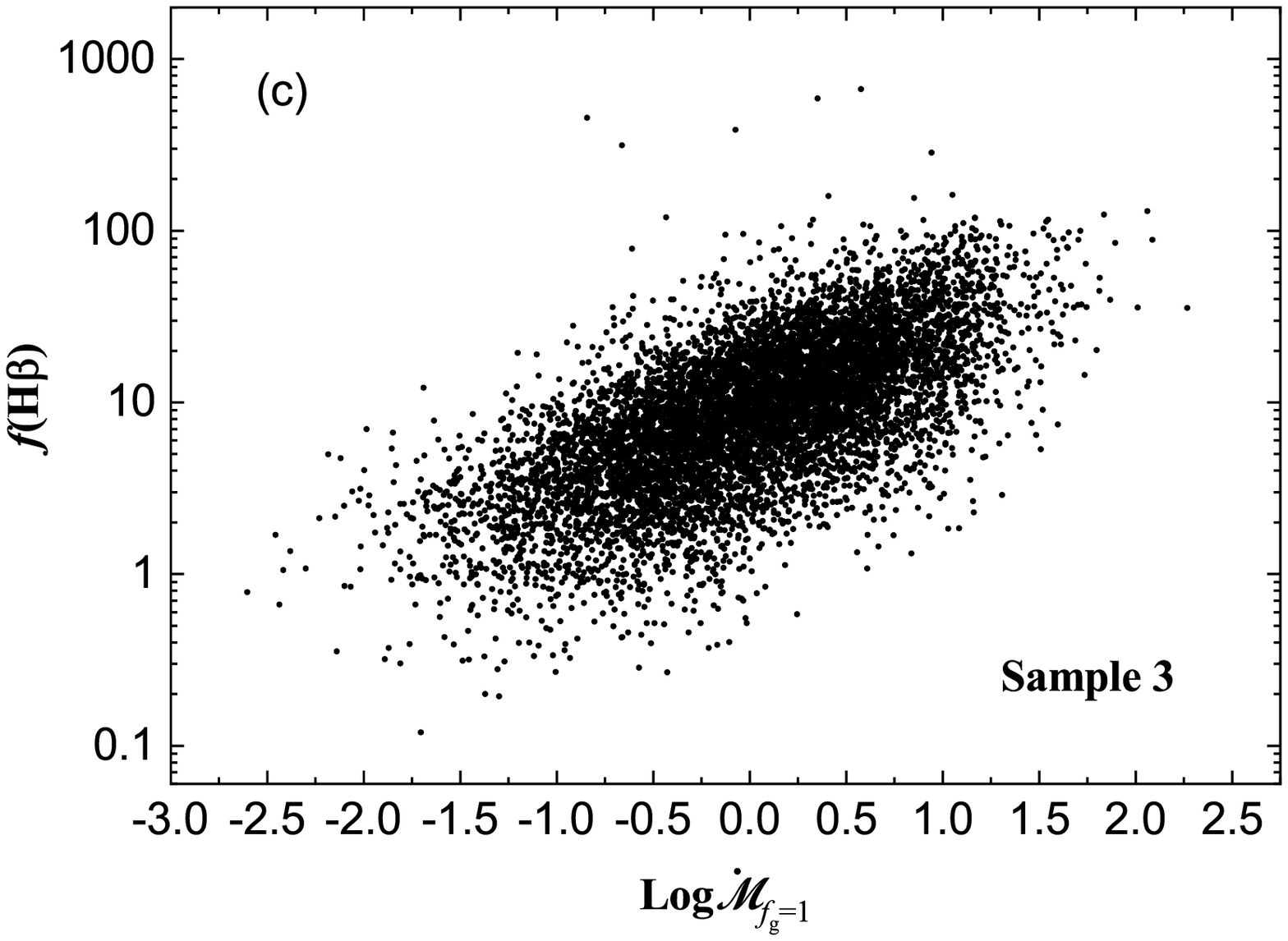}
   \includegraphics[angle=0,scale=0.26]{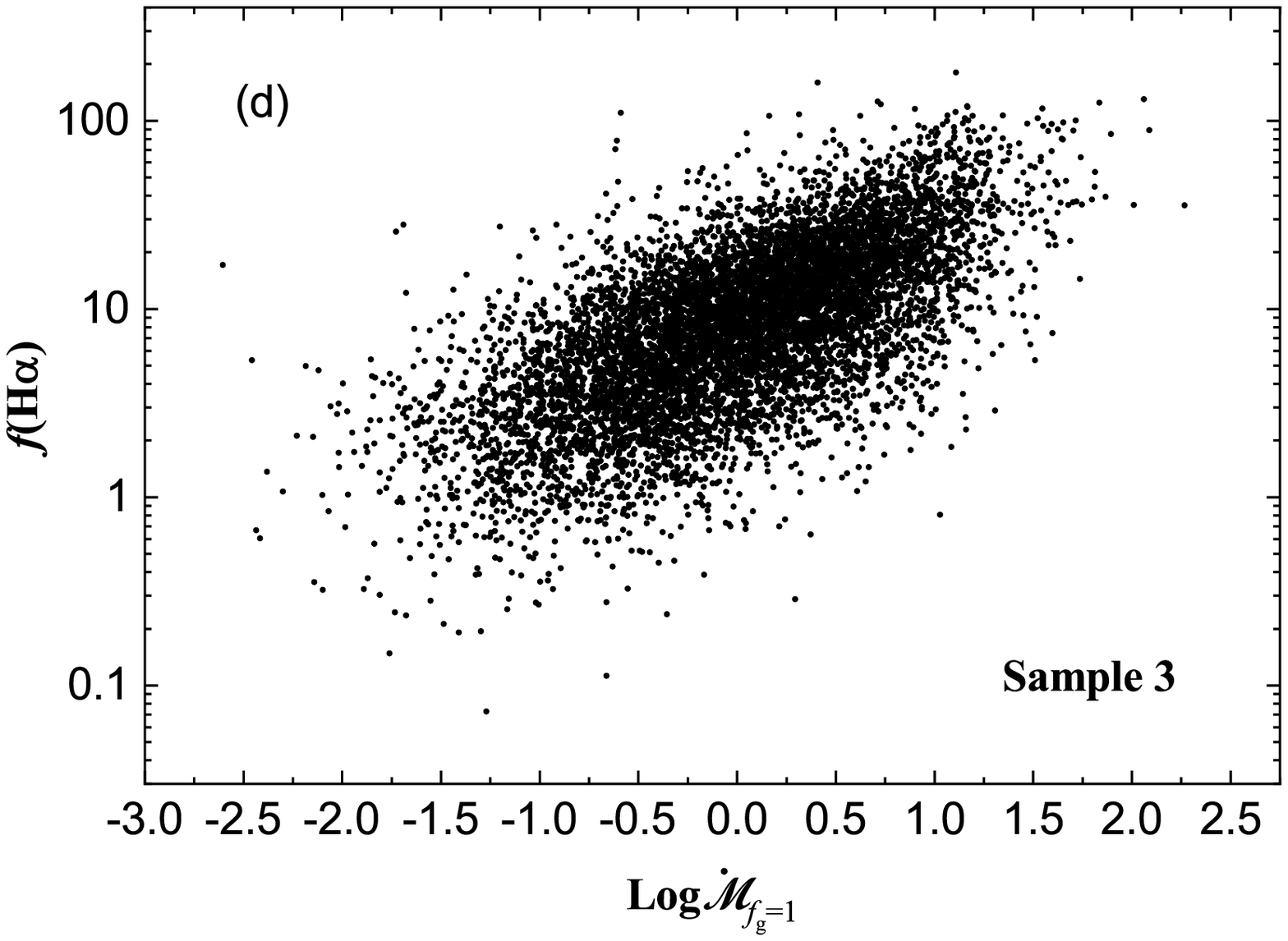}
  \end{center}
  \caption{$f$ vs. $\mathscr{\dot M}_{f_{\rm{g}}=1}$ for AGNs in Samples 1--3. The Spearman test shows positive correlations between these two physical quantities. $\mathscr{\dot M}_{f_{\rm{g}}=1}$ is estimated by $\eta=0.089M_{8}^{0.52}$ rather than $\eta=0.038$.}
  \label{fig:eta-f-rat}
\end{figure}

It is very difficult to get real individual value of $\eta$ to estimate $\mathscr{\dot M}_{f_{\rm{g}}=1}$ for a large sample of AGNs, because $\eta$ is closely related to the difficultly measured spin of a black hole. Usually, the Eddington ratio is regarded as a proxy of accretion rate of black hole. Even though these correlations of $\mathscr{\dot M}_{f_{\rm{g}}=1}$ with $f$ are likely influenced by the unknown individual value of $\eta$, there are still correlations of the Eddington ratio with $f$, because only a difference of 0.038 exists between $\mathscr{\dot M}_{f_{\rm{g}}=1}$ and $L_{\rm{bol}}/L_{\rm{Edd}}$ in Tables~\ref{sam1}--\ref{sam3}. \citet{Da11} found a strong correlation of $\eta=0.089 M_{8}^{0.52}$ for a sample of 80 Palomar--Green quasars, where $M_{8}$ is the black hole mass in units of $10^8 M_{\odot}$ and $\eta$ was estimated from the mass accretion rate and $L_{\rm{bol}}$. This empirical relation is used to estimate $\eta$ in order to test the influence of using $\eta=0.038$ on these correlations of $\mathscr{\dot M}_{f_{\rm{g}}=1}$ with $f$. Correlation analyses are made for $\mathscr{\dot M}_{f_{\rm{g}}=1}$ and $f$ in Figure~\ref{fig:fac-rat} with $\mathscr{\dot M}_{f_{\rm{g}}=1}$ to be re-estimated by $L_{\rm{bol}}/L_{\rm{Edd}}$ in Tables~\ref{sam1}--\ref{sam3} and the estimated $\eta$. There are still correlations very similar to those found in Figure~\ref{fig:fac-rat} when using these new dimensionless accretion rates (see Figure~\ref{fig:eta-f-rat} and Table~\ref{spearman}). Thus, these $\mathscr{\dot M}_{f_{\rm{g}}=1}$--$f$ correlations found in this work do not result from using the fixed value of $\eta=0.038$.

\begin{figure}
  \begin{center}
   \includegraphics[angle=0,scale=0.26]{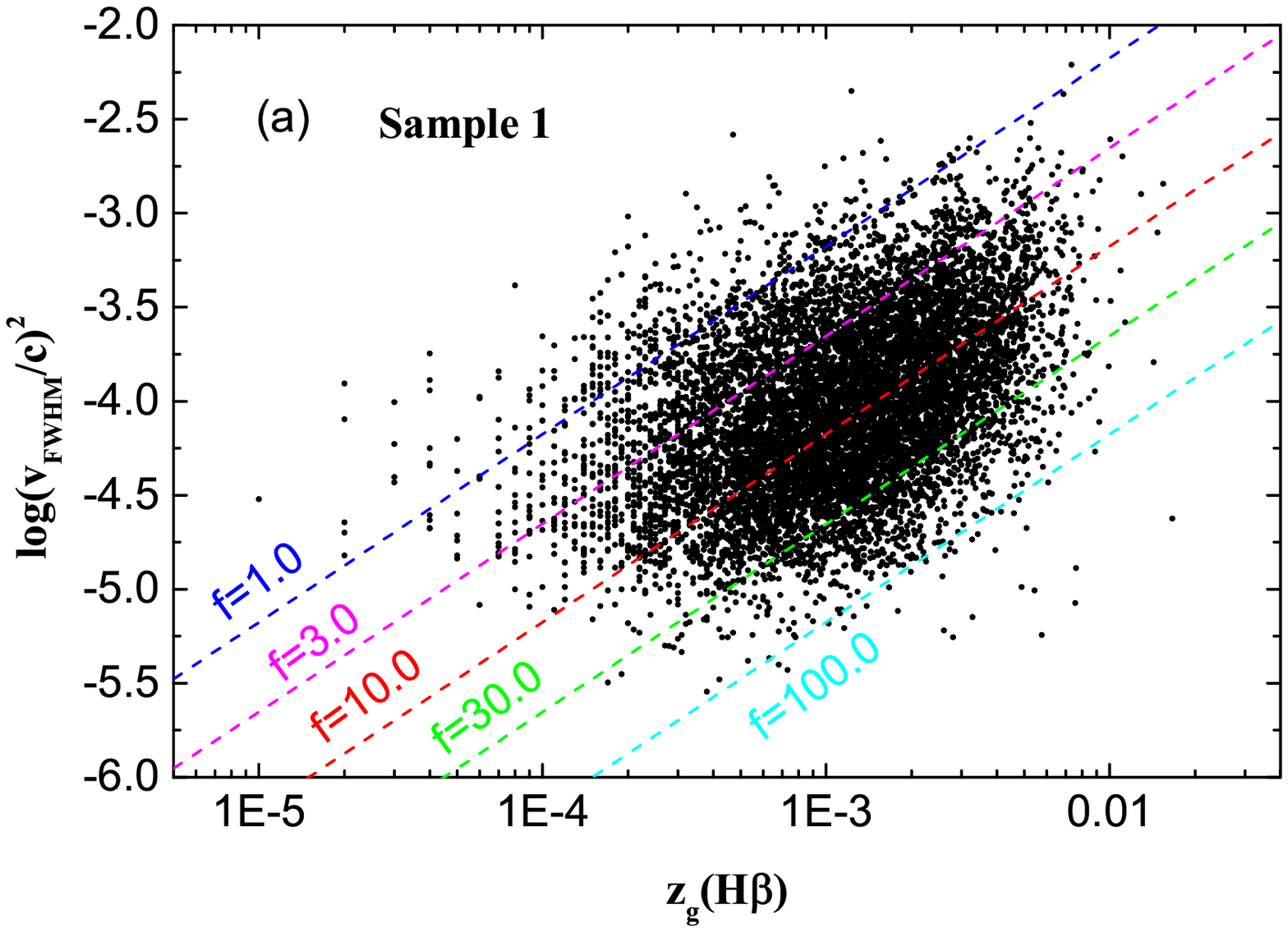}
   \includegraphics[angle=0,scale=0.26]{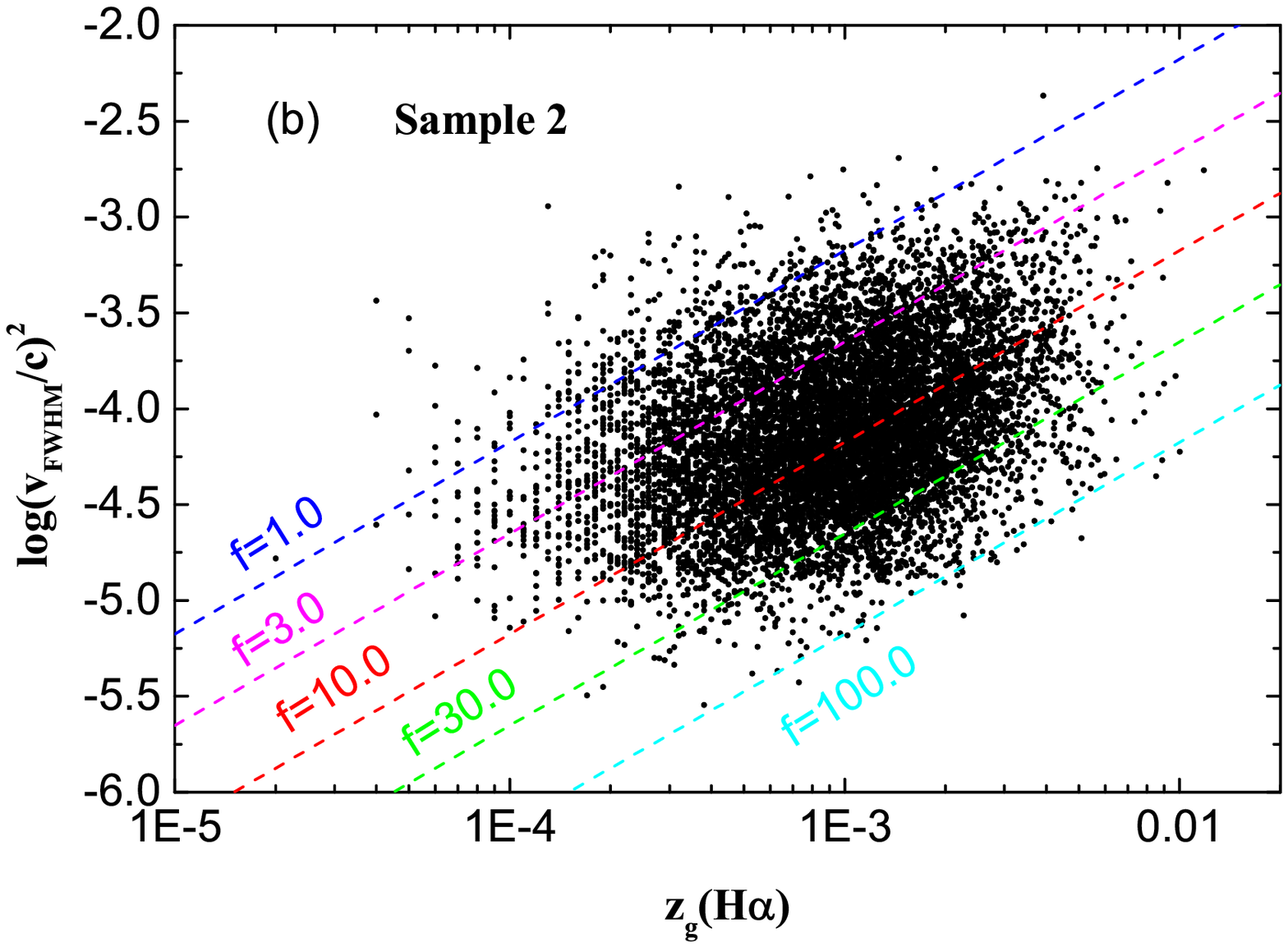}
   \includegraphics[angle=0,scale=0.26]{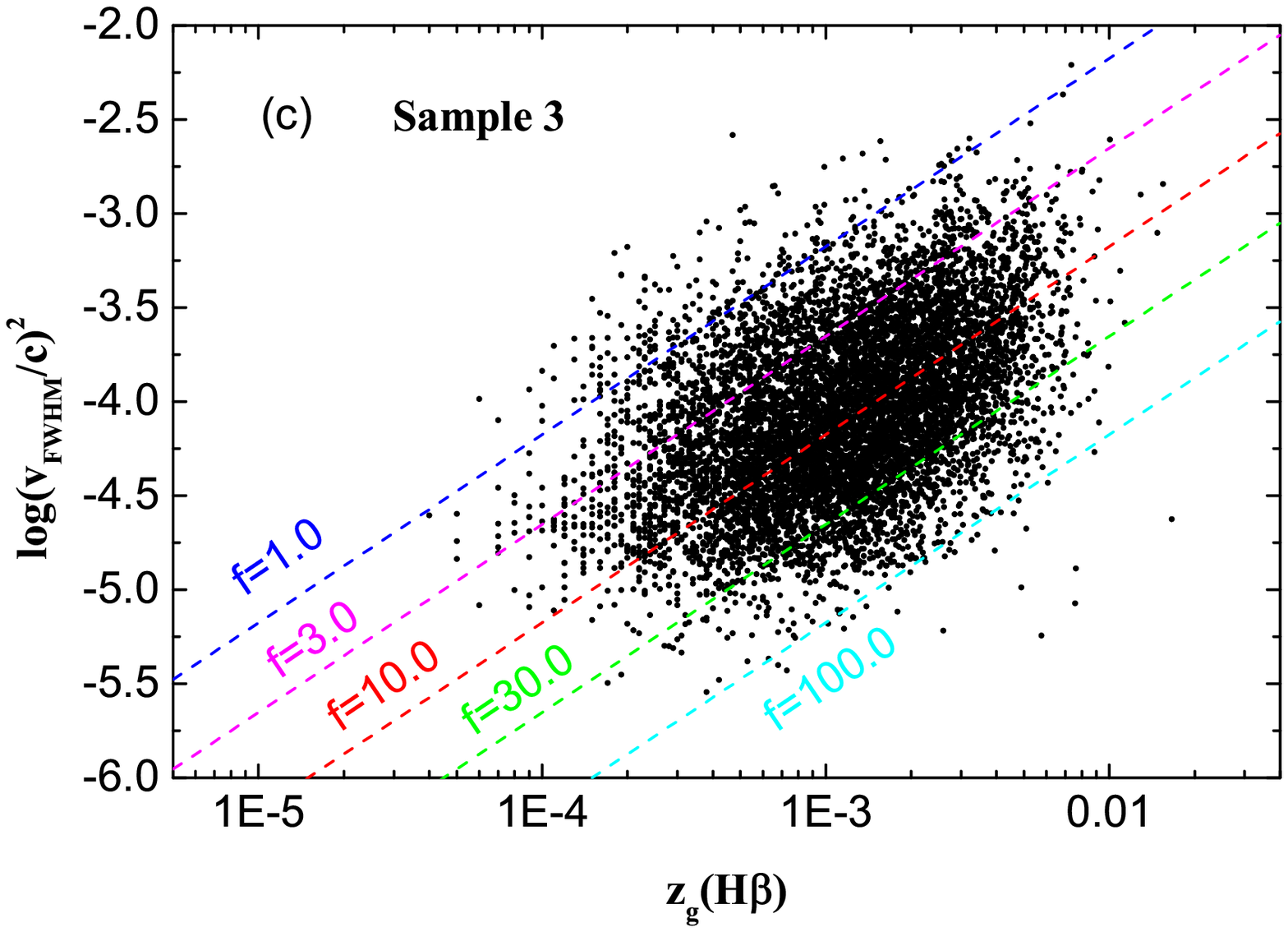}
   \includegraphics[angle=0,scale=0.26]{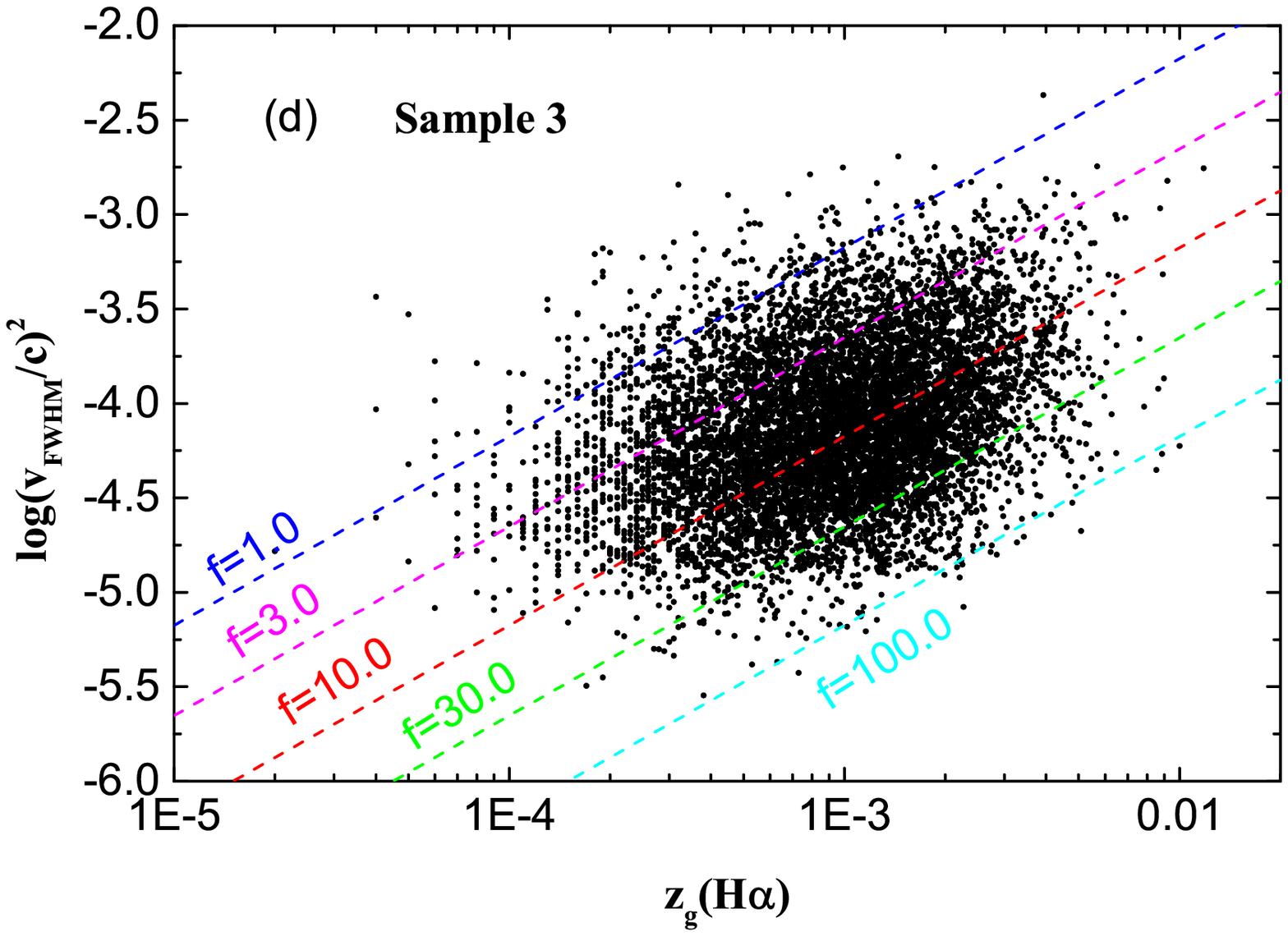}
  \end{center}
  \caption{$(v_{\rm{FWHM}}/c)^2$ vs. $z_{\rm{g}}$ for AGNs in Samples 1--3. The values labeled on the dashed lines represent $f$ in Equation (5).}
  \label{fig:fwhm-shif}
\end{figure}

Equation (2) can give for $v_{\rm{FWHM}}$, $f$, and $z_{\rm{g}}$
\begin{equation}
  \log (\frac{v_{\rm{FWHM}}}{c})^2 = -\log(\frac{3}{2}f)+\log z_{\rm{g}},
\end{equation}
which is similar to Equation (6) in \citet{MJ18}. \citet{MJ18} found a tight correlation between the widths and redward shifts of the Fe {\sc iii}$\lambda\lambda$ 2039--2113 blend for lensed quasars, which supports the gravitational interpretation of the Fe {\sc iii}$\lambda\lambda$ 2039--2113 redward shifts. A series of lines based on Equation (4) with different $f$ are compared to the observational data points (see Figure~\ref{fig:fwhm-shif}). From top to bottom, the corresponding $f$ increases. Because of the codependence among the Eddington ratio, dimensionless accretion rate and $v_{\rm{FWHM}}$, the large ranges of the former two quantities may lead to the large span in the direction roughly perpendicular to these lines (see Figure~\ref{fig:fwhm-shif}). These lines with $f=1$--100 recover the observational data in Figure~\ref{fig:fwhm-shif}, and this indicates the gravitational origin of $z_{\rm{g}}$. At the same time, the internal physical processes, e.g., the micro-turbulence, within the BLR cloud can broaden and smooth the line profles \citep{BF00}. Also, the turbulence velocity of the BLR cloud can influence the widths of the line profiles. These turbulence processes will influence $v_{\rm{FWHM}}$ and then $f$ for different AGNs. The combination of the column density of the BLR cloud, metallicity of the BLR cloud, internal physical processes within the BLR cloud, etc, may decrease these correlations in Figure~\ref{fig:fac-rat} \citep[e.g.,][]{Li22}.

There are the various outflows at accretion disk scales, the BLR scales, the NLR scales and
the kpc scales, driven by $F_{\rm{r}}$ from AGNs \citep{Ka18,Dy18,Da19,Ma19,No20,Me21,Si21}.
Thus, $F_{\rm{r}}$ is prevalent, and may contribute to the force budget for inflow, e.g., $F_{\rm{r}}$ decelerates inflow \citep{Fe09}. RM observations of PG 0026+129 indicate a decelerating inflow if $z_{\rm{g}}$ originates from inflow. If the decelerating inflow is prevalent, $z_{\rm{g}}$ will increase with the increasing $r_{\rm{BLR}}/r_{\rm{g}}$, but this expectation is not consistent with the negative trend found in Figure~\ref{fig:shif-rad}. Thus, the inflow seems not to be the origin of $z_{\rm{g}}$. In RM observations, the asymmetric lag maps and shifts of broad emission lines for AGNs usually differ from the theoretical expectation that inflow will generate the redward shifted broad emission lines with the blueward asymmetric lag maps \citep[e.g.,][]{De10,Zh19,Hu20,Fe21a,Fe21b}. This kind of broad emission lines may originate from an elliptical disklike BLR \citep{KW20,Fe21a}. Therefore, the redward shifted broad emission lines in AGNs do not necessarily originate from inflow.

\begin{figure}
  \begin{center}
   \includegraphics[angle=0,scale=0.26]{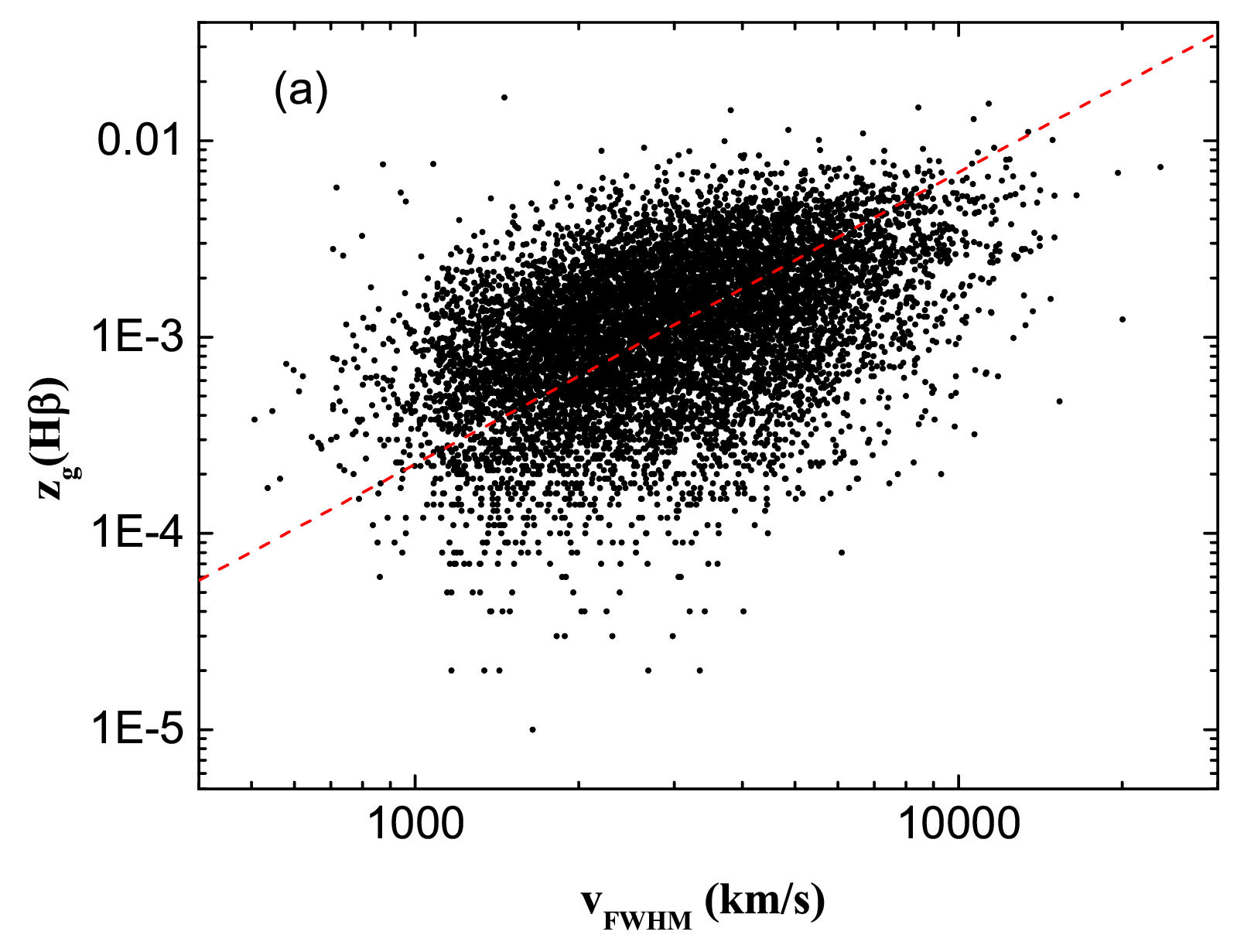}
   \includegraphics[angle=0,scale=0.26]{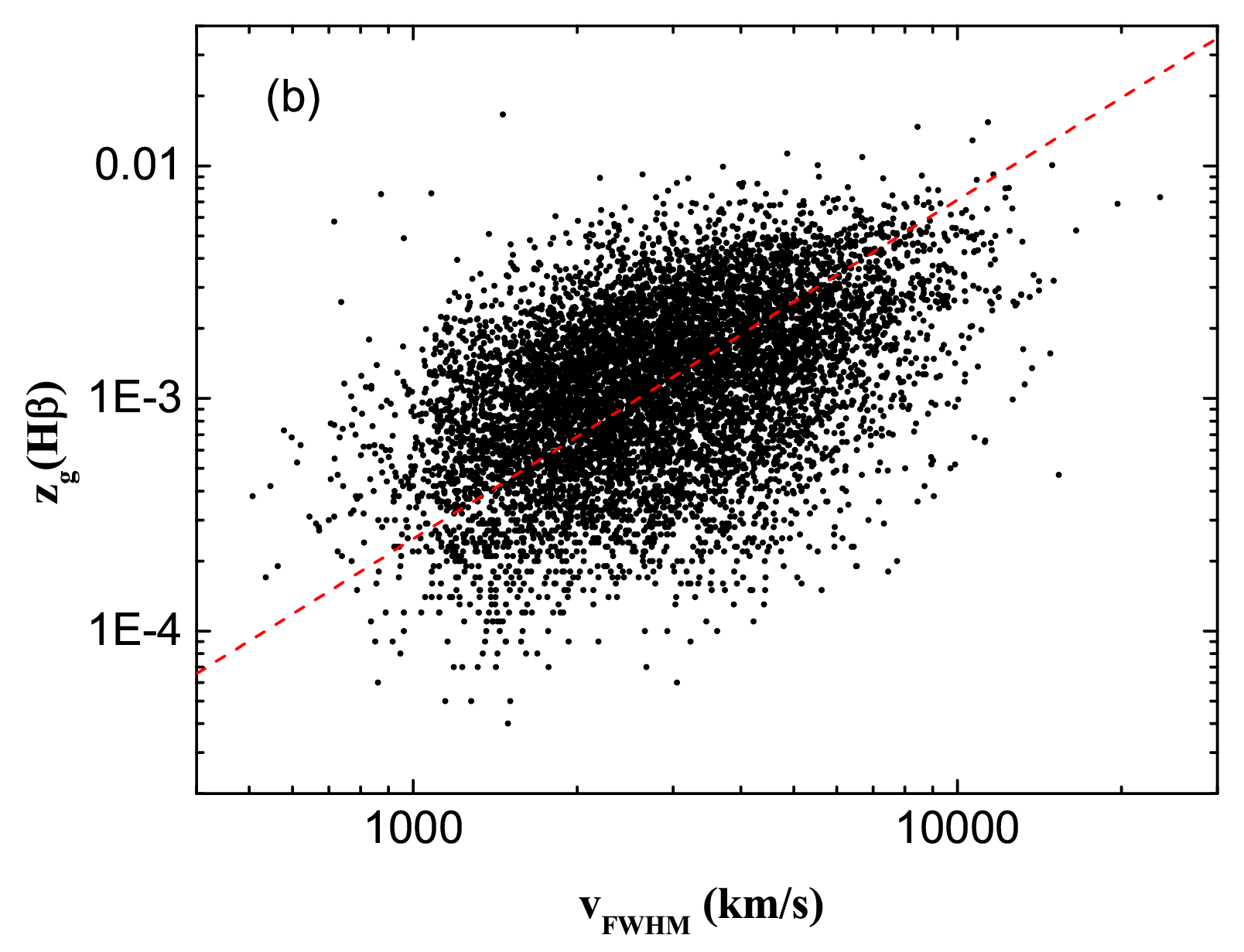}
  \end{center}
  \caption{Panel ($a$): $z_{\rm{g}}$ vs. $v_{\rm{FWHM}}$ for \Hb\ in Sample 1. The blue dashed line represents the best bisector fit, $\log z_{\rm{g}} = -8.106(\pm 0.073)+1.486(\pm 0.021)\log v_{\rm{FWHM}}$, with the $p$-value of the hypothesis test to be $<10^{-40}$. Panel ($b$): $z_{\rm{g}}$ vs. $v_{\rm{FWHM}}$ for \Hb\ in Sample 3. The red dashed line represents the best bisector fit, $\log z_{\rm{g}} = -7.978(\pm 0.072)+1.458(\pm 0.021)\log v_{\rm{FWHM}}$, with the $p$-value of the hypothesis test to be $<10^{-40}$.}
  \label{fig:redwid}
\end{figure}

\citet{MR18} determined the virial factor in a smaller set of sources using a different method than proposed here, and found a relation whereby $f \propto 1/v_{\rm{FWHM}}$, which is attributed to inclination effects, but without excluding the possibility of radiation pressure effects over a wide luminosity range. Their sources have $\log [v_{\rm{FWHM}}/\rm{(km~ s^{-1}})]$ $\approx$ 3.2--4.0, which are much narrower than $\log [v_{\rm{FWHM}}/\rm{(km~ s^{-1}})]$ $\approx$ 2.7--4.4 in our samples. Also, their sources have $\log (M_{\rm{RM}}/M_{\odot})\approx$ 7.5--9.7 and $\log [L_{5100}/\rm{(erg~s^{-1}})]$ $=$ 44.3--46.2, which are much narrower than $\log (M_{\rm{RM}}/M_{\odot})$ $\approx$ 5.2--9.7 and $\log [L_{5100}/\rm{(erg~s^{-1}})]$ $=$ 40.6--45.6 in our samples, respectively. There are positive correlations between $z_{\rm{g}}$ and $v_{\rm{FWHM}}$ for our samples, $z_{\rm{g}}\propto v_{\rm{FWHM}}^{1.5}$ (see Figure~\ref{fig:redwid}). Based on $z_{\rm{g}}\propto v_{\rm{FWHM}}^{1.5}$ and Equation (2) with $v_{\rm{FWHM}}$ partly contributed from inclination effects, we have $f\propto 1/v_{\rm{FWHM}}^{0.5}$, which is qualitatively consistent with, but shallower than $f \propto 1/v_{\rm{FWHM}}$. This discrepancy might be generated by our consideration of radiation pressure, and the estimation of $M_{\bullet}$ using standard thin accretion disk models for sources with the narrower parameter coverage \citep{MR18}. In this sense, \textbf{these} results and interpretations promoted here are consistent with \citet{MR18}.

\begin{figure}
  \begin{center}
   \includegraphics[angle=0,scale=0.26]{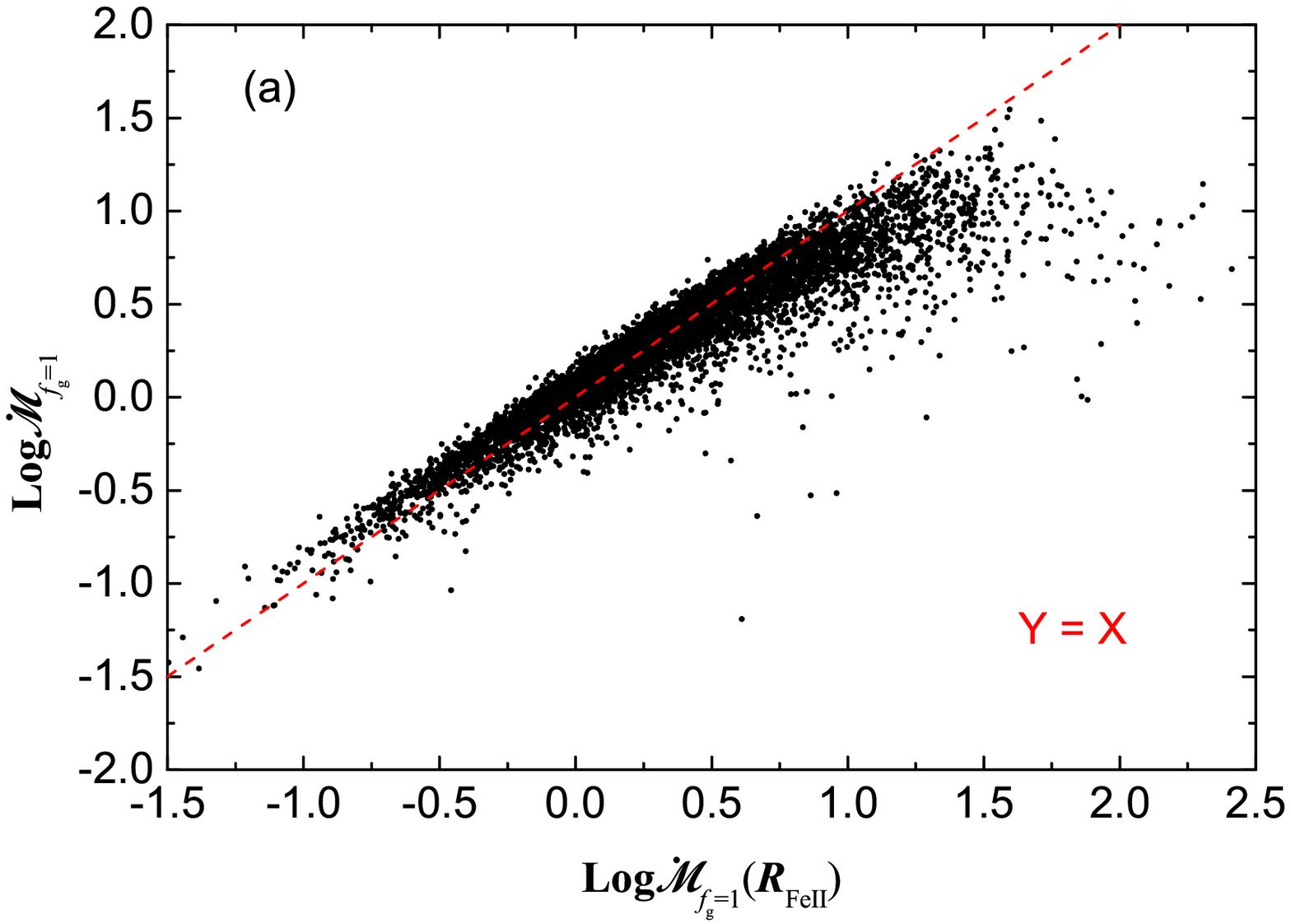}
   \includegraphics[angle=0,scale=0.26]{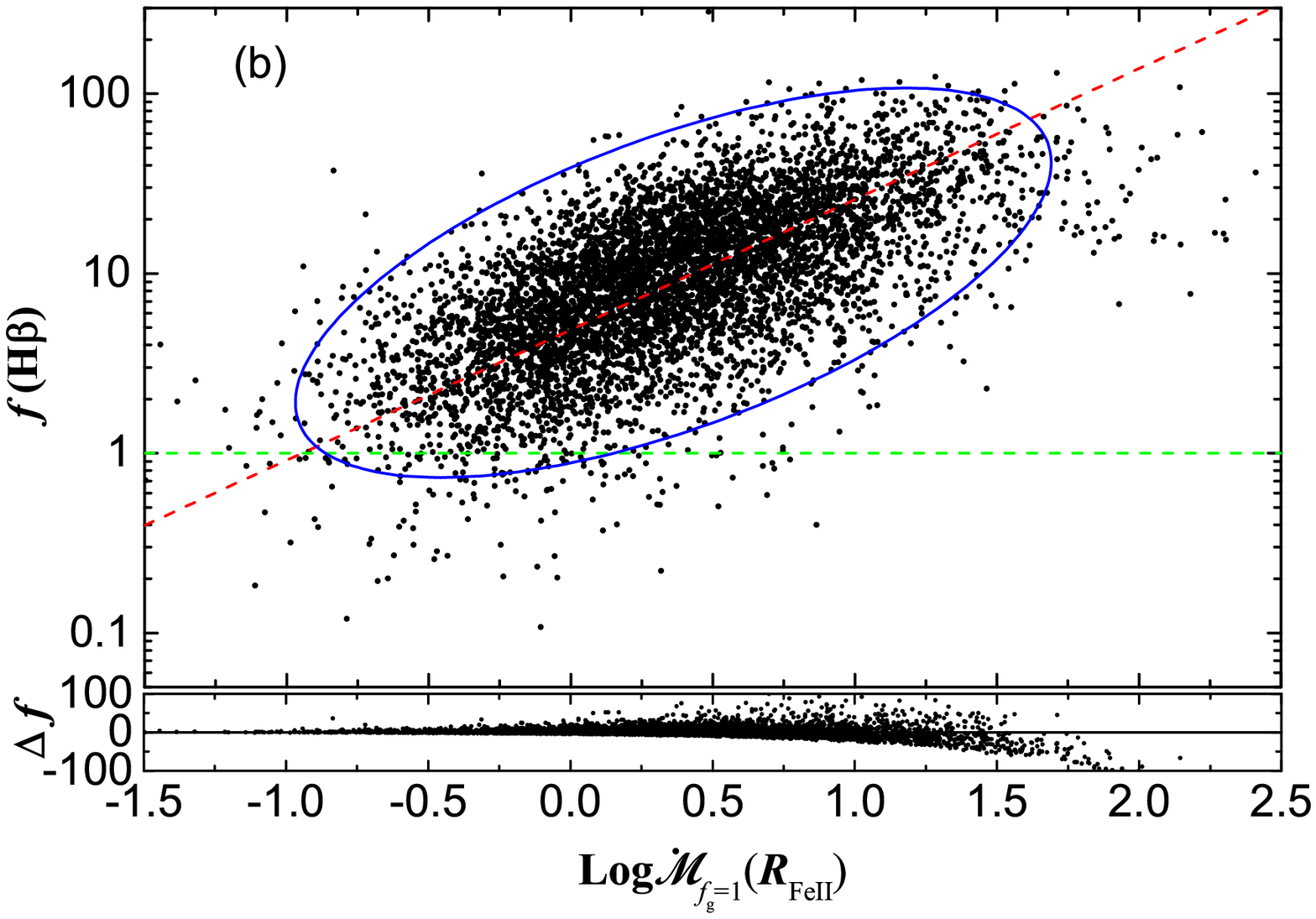}
   \includegraphics[angle=0,scale=0.26]{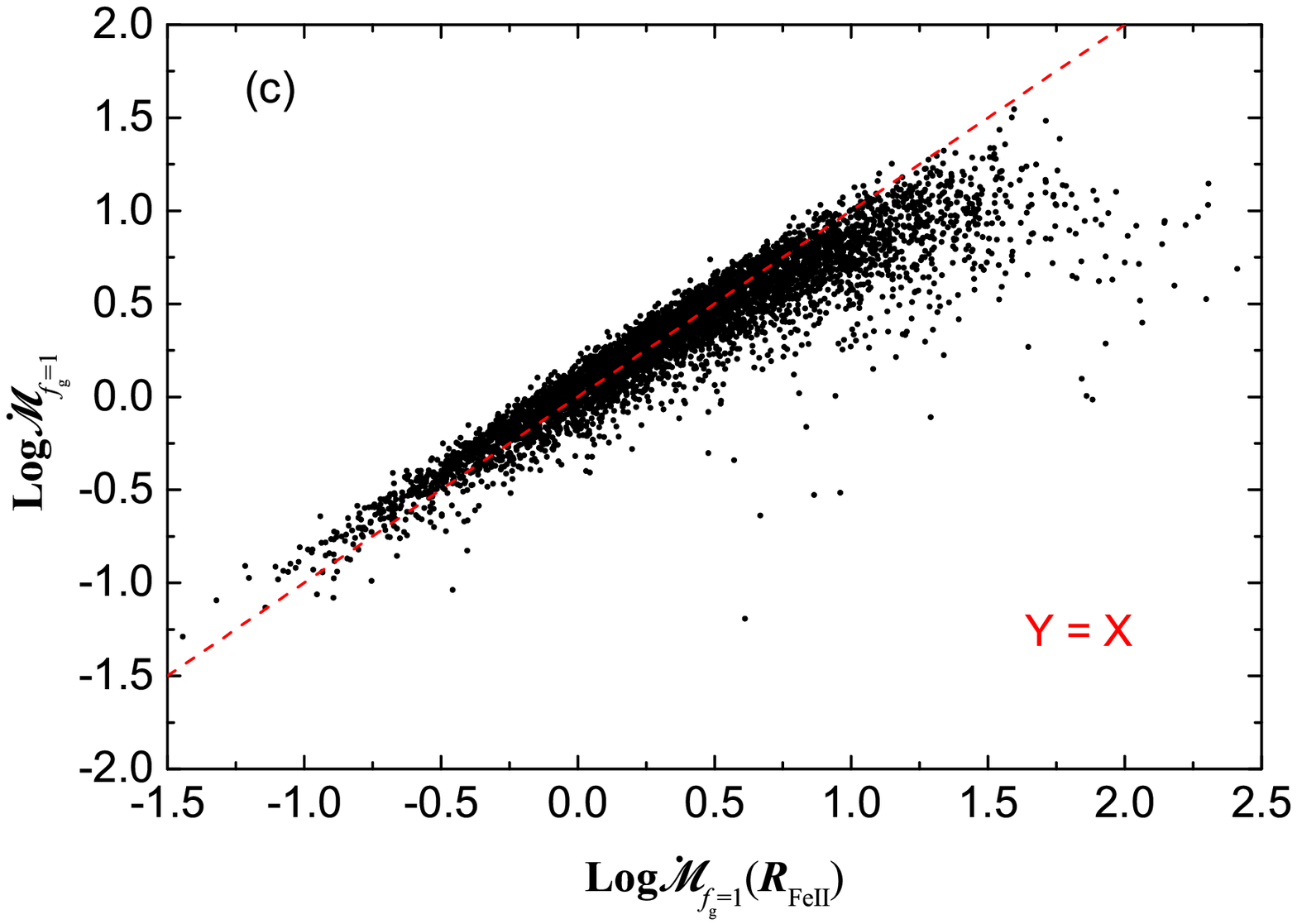}
   \includegraphics[angle=0,scale=0.26]{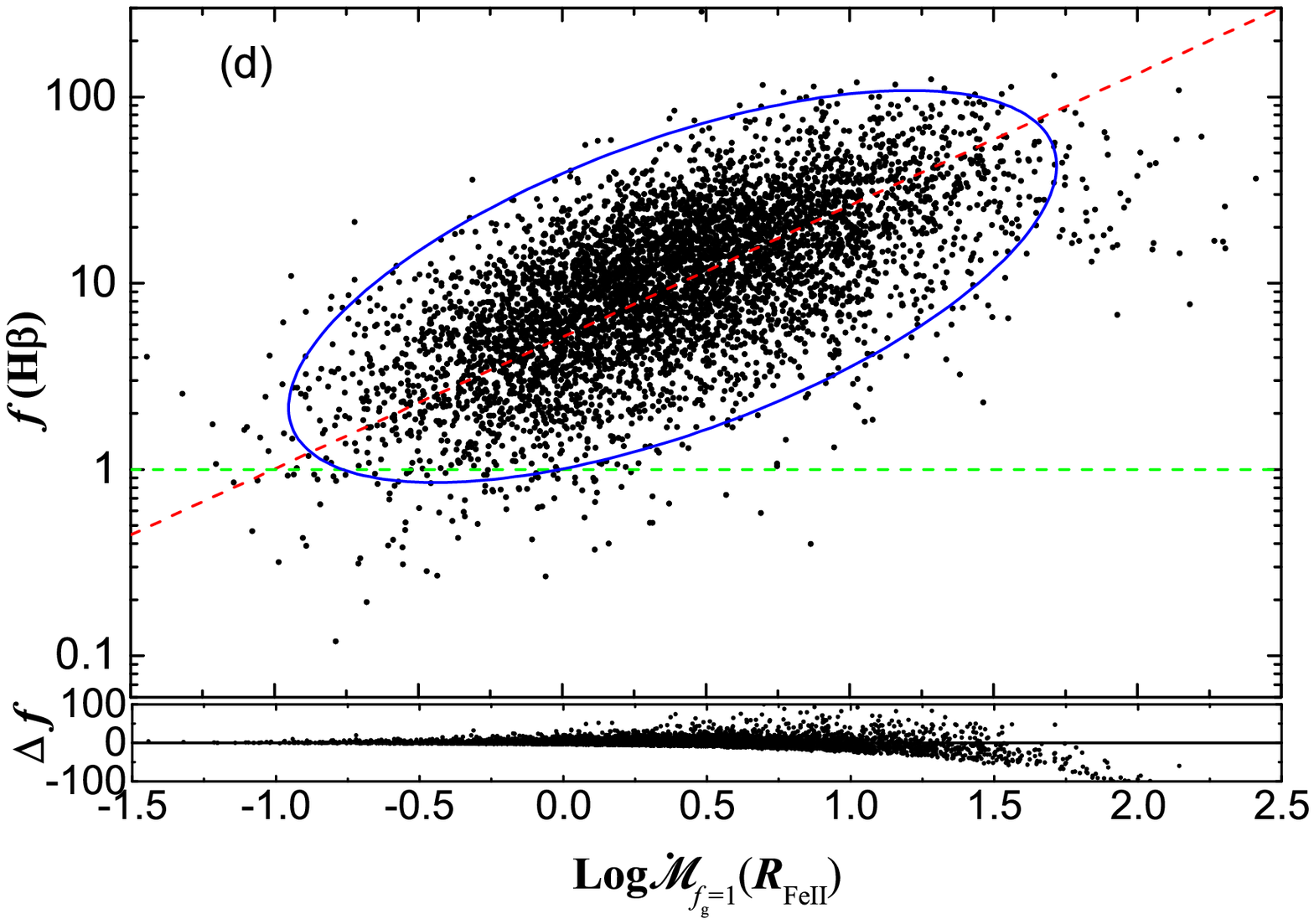}
  \end{center}
  \caption{Panel($a$): $\mathscr{\dot M}_{f_{\rm{g}}=1}$ vs. $\mathscr{\dot M}_{f_{\rm{g}}=1}(R_{\rm{FeII}})$ for \Hb\ in 5997 AGNs from Sample 1. Panel($b$): $f$ vs. $\mathscr{\dot M}_{f_{\rm{g}}=1}(R_{\rm{FeII}})$ for \Hb\ in 5997 AGNs from Sample 1. The best bisector fit is $\log f = 0.69(\pm0.01)+0.73(\pm0.01) \log \mathscr{\dot M}_{f_{\rm{g}}=1}(R_{\rm{FeII}})$, with the $p$-value of the hypothesis test to be $<10^{-40}$. Panel($c$): $\mathscr{\dot M}_{f_{\rm{g}}=1}$ vs. $\mathscr{\dot M}_{f_{\rm{g}}=1}(R_{\rm{FeII}})$ for \Hb\ in 5365 AGNs from Sample 3. Panel($d$): $f$ vs. $\mathscr{\dot M}_{f_{\rm{g}}=1}(R_{\rm{FeII}})$ for \Hb\ in 5365 AGNs from Sample 3. The best bisector fit is $\log f = 0.71(\pm0.01)+0.71(\pm0.01) \log \mathscr{\dot M}_{f_{\rm{g}}=1}(R_{\rm{FeII}})$, with the $p$-value of the hypothesis test to be $<10^{-40}$. The coloured lines in Panels ($b$) and ($d$) are same as in Figure~\ref{fig:fac-rat}, and two outliers with $\log \mathscr{\dot M}_{f_{\rm{g}}=1}(R_{\rm{FeII}})\approx$ 27 and 103 are not included in fitting.}
  \label{fig:acc-rat}
\end{figure}

\begin{figure}
  \begin{center}
   \includegraphics[angle=0,scale=0.26]{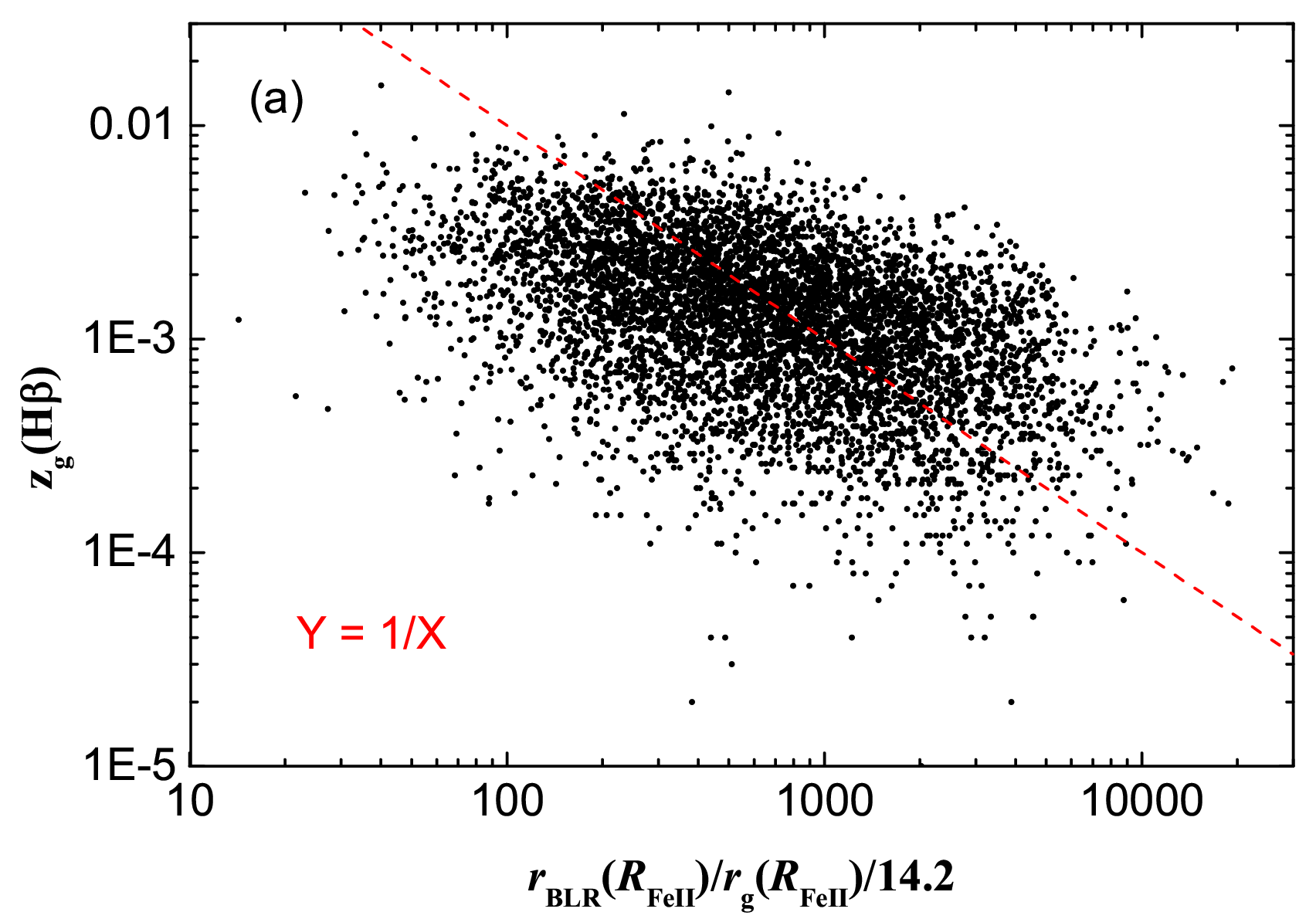}
   \includegraphics[angle=0,scale=0.26]{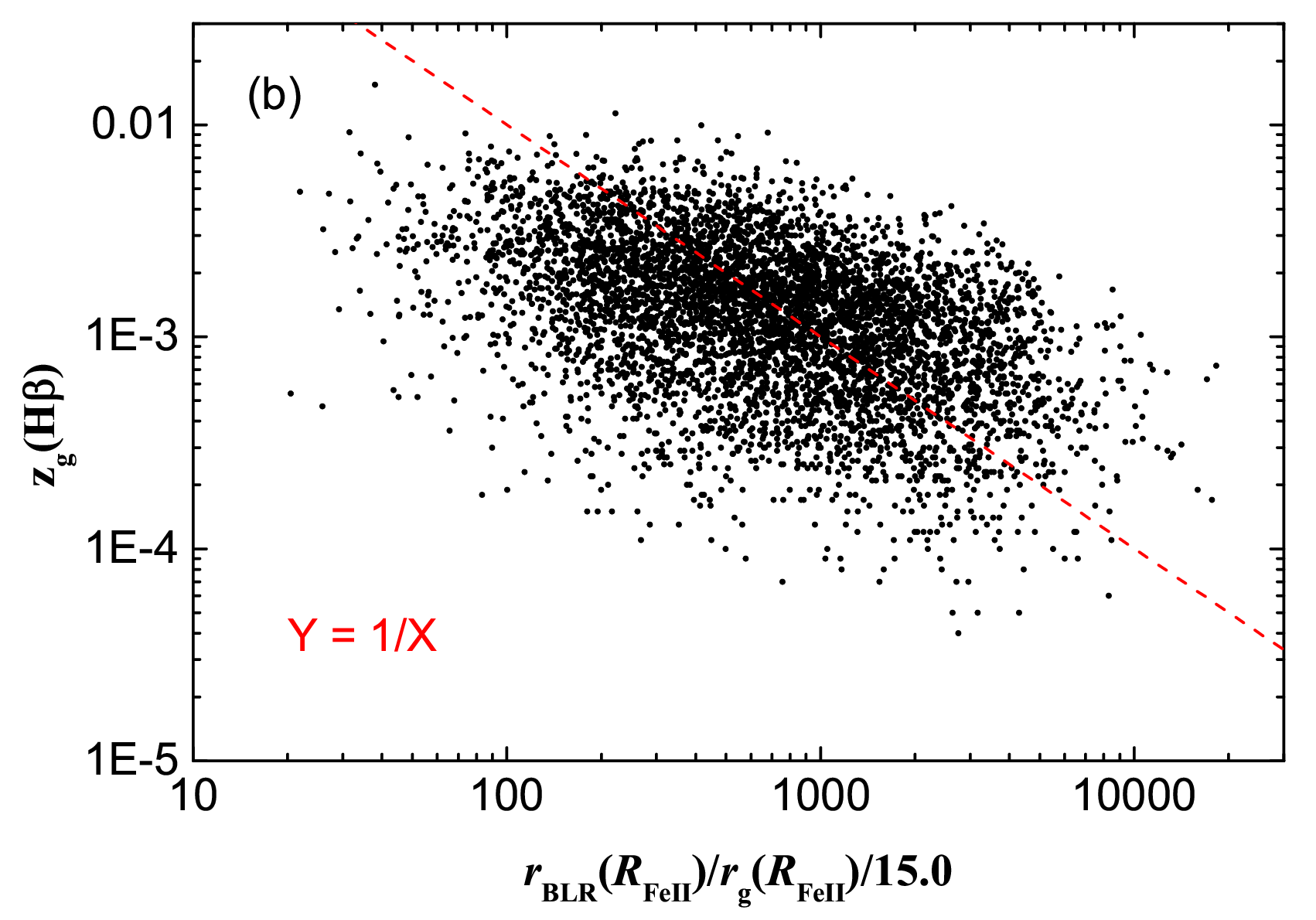}
  \end{center}
  \caption{Panel($a$): \Hb\ shift $z_{\rm{g}}$ vs. $r_{\rm{BLR}}(R_{\rm{FeII}})/r_{\rm{g}}(R_{\rm{FeII}})$ corrected by $\langle f \rangle =14.2$ for 5997 AGNs in Sample 1. Panel($b$): \Hb\ shift $z_{\rm{g}}$ vs. $r_{\rm{BLR}}(R_{\rm{FeII}})/r_{\rm{g}}(R_{\rm{FeII}})$ corrected by $\langle f \rangle =15.0$ for 5365 AGNs in Sample 3.}
  \label{fig:shift-radius}
\end{figure}

The AGNs with high-accretion rates show shorter time lags by factors of a few compared to the predictions from the $r_{\rm{BLR}}$--$L_{\rm{5100}}$ relationship \citep{Du15}. \citet{Du19} found that accretion rate is the main driver for the shortened lags, and established a new scaling relation:
\begin{equation}
\log r_{\rm{BLR}}(\mathrm{H\beta}) = 1.65+0.45\log L_{44}- 0.35 R_{\rm{FeII}},
\end{equation}
where $r_{\rm{BLR}}(\rm{H\beta})$ is $r_{\rm{BLR}}$ in units of light days for \Hb, $L_{44}=L_{\rm{5100}}/(\rm{10^{44}~erg~s^{-1}})$, and $R_{\rm{FeII}}$ is the line ratio of \feii\ to \Hb. Replacing $r_{\rm{BLR}}=33.65 L_{44}^{0.533}$ with Equation (6), the mass of black hole is given by
\begin{equation}
  \log M_{\rm{RM}}(R_{\rm{FeII}})=\log M_{\rm{RM}}-0.083\log L_{44}-0.35 R_{\rm{FeII}}+0.123,
\end{equation}
which is used to estimate the dimensionless accretion rate, $\mathscr{\dot M}_{f_{\rm{g}}=1}(R_{\rm{FeII}})$. Samples 1 and 3 are used to investigate the influence of Equation (6) on the $f$--$\mathscr{\dot M}_{f_{\rm{g}}=1}$ relation. $R_{\rm{FeII}}$ is estimated by equivalent widths of \Hb\ and \feii\ taken from Table 2 of \citet{Li19} for 5997 AGNs in Sample 1 and 5365 AGNs in Sample 3. First, $\mathscr{\dot M}_{f_{\rm{g}}=1}(R_{\rm{FeII}})$ is overall consistent with the original $\mathscr{\dot M}_{f_{\rm{g}}=1}$ (see Figure~\ref{fig:acc-rat}). Second, $f$ is well correlated with $\mathscr{\dot M}_{f_{\rm{g}}=1}(R_{\rm{FeII}})$ (see Figure~\ref{fig:acc-rat}), and Equation (6) has a slight impact on the $f$--$\mathscr{\dot M}_{f_{\rm{g}}=1}$ relation. Also, $r_{\rm{BLR}}(R_{\rm{FeII}})/r_{\rm{g}}(R_{\rm{FeII}})$ is estimated, and there exits the anti-correlation trend between $z_{\rm{g}}$ and $r_{\rm{BLR}}(R_{\rm{FeII}})/r_{\rm{g}}(R_{\rm{FeII}})/\langle f \rangle$ (see Figure~\ref{fig:shift-radius}), same as in Figure~\ref{fig:shif-rad}. The potential effect of $\mathscr{\dot M}$, especially at the high mass accretion rate end \citep{Du15}, do not lead to qualitatively different results of $r_{\rm{BLR}}/r_{\rm{g}}$.

\section{CONCLUSION}\label{sec:conc}
Based on the assumption of a gravitational origin for the redward shifts of broad emission lines \Hb\ and \Ha, and their widths and redward shifts for more than 8000 SDSS DR7 AGNs with $z<0.35$, we measured the virial factor in $M_{\rm{RM}}$, estimated by the RM method and/or the relevant secondary methods. The measured virial factor contains the overall effect of $F_{\rm{r}}$ from accretion disk radiation and the geometric effect of BLR. Our findings can be summarized as follows:
\begin{enumerate}
  \item There are positive correlations of $f$ with $\mathscr{\dot M}_{f_{\rm{g}}=1}$ and $L_{\rm{bol}}/L_{\rm{Edd}}$, which are a combined effect of several physical mechanisms, such as the Doppler effects, the gravitational redshift, the gravity of black hole, the radiation pressure force, etc. $f$ spans a large range, and $f>1$ for $>$96\% AGNs in Samples 1--3. The $f$ correction makes the percent of high-accreting AGNs decrease by about 100 times, and blurs the distinction between high- and low-accreting sources.

  \item $z_{\rm{g}}$ is anti-correlated with $r_{\rm{BLR}}/r_{\rm{g}}$. $z_{\rm{g}}$ and $r_{\rm{BLR}}/r_{\rm{g}}/\langle f\rangle$ marginally follow the 1:1 line. A series of lines with different $f$ basically reproduce the $v_{\rm{FWHM}}$--$z_{\rm{g}}$ distribution for the broad \Hb\ and \Ha. These results suggest that the redward shifts of the broad \Hb\ and \Ha\ are governed by the gravity of the central SMBHs.

  \item For quasars at $z\gtrsim 6$, the $f$ correction makes them from the close Eddington accreting sources become low-accreting sources, likely in the radiatively efficient regime via a geometrically thin, optically thick accretion disk. The $f$ corrected masses indicate that quasars at $z\gtrsim 6$ have more massive early black hole seeds and longer growth times, supporting heavy-seed origin scenarios of early SMBHs. These results will make it more challenging to explain the formation and growth of SMBHs at $z\gtrsim 6$.

  \item 62 AGNs and 88 quasars, beyond the local Universe, do not follow these local $M_{\bullet}-\sigma_{\ast}$ relations. After the $f$ correction, these 150 sources are above these local $M_{\bullet}-\sigma_{\ast}$ relations, but they roughly follow the $f$-corrected $M_{\bullet}-\sigma_{\ast}$ relation of these local luminous AGNs in \citet{Ca20}. These results might shed light on possible redshift evolution in the $M_{\bullet}-\sigma_{\ast}$ relationship.
  \end{enumerate}

Our results show that radiation pressure force should be considered in estimating the virial masses of SMBHs. The usually used values of $f$ should be corrected for high-accreting AGNs, especially quasars at $z\gtrsim 6$. The $f$ correction to $M_{\rm{RM}}$ will make the coevolution (or not) of SMBHs and host galaxies more complex for the local sources and the higher redshift sources. Positive correlations of $f$ with $\mathscr{\dot M}_{f_{\rm{g}}=1}$ and $L_{\rm{bol}}/L_{\rm{Edd}}$ need to be further tested by the redward shifted broad emission lines of the RM AGNs without the signatures of inflow and outflow in BLR, which can be picked out by the velocity-resolved time lag maps.

\acknowledgements {We are grateful to the anonymous referee for constructive comments
and suggestions that improved significantly this manuscript. We thank the financial support of the National Key R\&D Program of China (grant No. 2021YFA1600404), the National Natural Science Foundation of China (grants No. 12373018, No. 12303022, No. 12203096, No. 12063005, and No. 11991051), Yunnan Fundamental Research Projects (grants No. 202301AT070358 and No. 202301AT070339), Yunnan Postdoctoral Research Foundation Funding Project, Special Research Assistant Funding Project of Chinese Academy of Sciences, and the science research grants from the China Manned Space Project with grant No. CMS-CSST-2021-A06. We acknowledge the Program for Innovative Research Team (in Science and Technology) in University of Yunnan Province (IRTSTYN).}

\begin{center}
\textbf{ORCID iDs}
\end{center}
H. T. Liu https://orcid.org/0000-0002-2153-3688 \\
Hai-Cheng Feng https://orcid.org/0000-0002-1530-2680\\
Sha-Sha Li https://orcid.org/0000-0003-3823-3419\\
H. Z. Li https://orcid.org/0000-0001-8307-1442

\label{lastpage}

\end{document}